\documentclass[aps,notitlepage,amsmath,amssymb,twocolumn]{revtex4-1}
\usepackage{graphicx}
\usepackage{hyperref}
\pdfoutput=1

\begin{document}

\title{Large scale spatio-temporal behaviour in surface growth}

\author{Vaidas Juknevi\v{c}ius}
\author{Julius Ruseckas}
\author{Jogundas Armaitis}

\affiliation{Institute of Theoretical Physics and Astronomy, Vilnius University,
Sauletekio~al.~3, LT-10222 Vilnius, Lithuania}

\begin{abstract}
 This paper presents new findings concerning the dynamics of the slow height variations in surfaces produced by the two-dimensional isotropic Kuramoto-Sivashinsky equation with an additional nonlinear term. 
In addition to the disordered patterns of specific size evident at small scales, slow height variations of scale-free character become increasingly evident when the system size is increased. The surface spectrum at small wave numbers has a power-law shape with a lower cut-off due to the finite system size. 
The temporal properties of these long-range height variations are investigated by analysing the time series of surface roughness fluctuations. The resulting power-spectral densities can be expressed as a sum of white noise and a generalized Lorentzian whose cut-off frequency varies with system size. 
The dependence of this lower cut-off frequency on the smallest wave number connects spatial and temporal properties and gives new insight into the surface evolution on large scales. 
\end{abstract}
\maketitle

\section*{Introduction}

%Surface growth is important
%It is described by nonlinear equations
%Not all parameter values accessible with current tech
%Still important having in mind progress

Detailed understanding of surface growth physics
has fueled several advances in science and technology,
including more accurate dating in archeology 
\cite{harris2014principles},
better integrated circuit technology \cite{mayfundamentals},
as well as production of novel materials \cite{bae2010roll}.
In many cases of scientific and technological interest, 
the evolution of growing surfaces can be described by so-called \emph{continuum models} that consist of nonlinear partial differential
equations and often display rich and interesting dynamics 
\cite{RevModPhys.65.851}.
Even though not all of this dynamics is currently accessible 
experimentally, it still is worthwhile to investigate,
especially in the view of rapid experimental \cite{castro12} and 
theoretical \cite{hairer13} progress.

The object of this study is a continuum surface growth model described by the two-dimensional \emph{generalized} Kuramoto-Sivashisky equation with a single independent parameter $\alpha$,
\begin{equation}
	\partial_t h = -\nabla^2 h - \nabla^4 h- \alpha \nabla^2(\nabla h)^2 + (\nabla h)^2\,,
	\label{eq:gkse}
\end{equation}
considered in \cite{juknevicius2016long}, that produces chaotically evolving disordered spatial patterns. 
Equations of this type (with and without added noise) have been successfully used as models for amorphous solid surface growth  \cite{raible2000amorphous2,raible2001amorphous,raible2002growth} and nano-scale pattern formation induced by ion beam sputtering (IBS) \cite{{cuerno1995dynamic},{kim2004kinetic},castro2005self,gago2006order,cuerno2011nanoscale}.

Eq.~(\ref{eq:gkse}) in two spatial dimensions describes the evolution of a (2+1)-dimensional interface, i.e., a surface whose height $h(\boldsymbol{r},t)$ is defined as a function on a two-dimensional plane $\boldsymbol{r}\in\mathbb{R}^2$ that is growing in the direction $h$ perpendicular to that plane as time $t$ goes by.
Numerical studies of Eq.~(\ref{eq:gkse}) in one dimension have also been performed by Mu{\~n}oz-Garcia et. al.~\cite{munoz2006short}, and a good correspondence to the IBS experiments has been found \cite{munoz2010observation}.

Eq.~(\ref{eq:gkse}) has the celebrated Kuramoto-Sivashinsky (KS) equation \cite{sivashinsky1977nonlinear,michelson1977nonlinear,sivashinsky1979self,
kuramoto1984chemical} as its special case when parameter $\alpha=0$:
\begin{equation}
	\partial_t h = -\nabla^2 h - \nabla^4 h + (\nabla h)^2\,.
	\label{eq:kse}
\end{equation}
The latter equation stands as a paradigmatic model for chaotic spatially extended
systems and has been used to study the connections between chaotic dynamics at
small scales and apparent stochastic behaviour at large scales 
\cite{sneppen1992dynamic,jayaprakash1993universal,
boghosian1999hydrodynamics}. 
Various generalizations and modifications of the KS equation (\ref{eq:kse}) with local and non-local damping terms, anisotropy, and noise have been used to study pattern formation due to ion-beam erosion \cite{paniconi1997stationary,rost1995anisotropic,
lauritsen1996noisy,dreimann2010continuum,diddens2013continuum,diddens2015continuum}.
Eq.~(\ref{eq:kse}) itself in
one- and two-dimensional cases has been a subject of active research for about
three decades, and its scaling properties have even been an object of
controversy.

%%%%%%%%%%%%%%%%%%%%%%%%%%%%
It has been suggested by Yakhot \cite{yakhot1981large} and subsequently confirmed and reiterated by different authors (see e.g., \cite{sneppen1992dynamic,procaccia1992surface,jayaprakash1993universal}) that the large-scale behaviour of the deterministic KS equation (\ref{eq:kse}) in the one-dimensional case can be described by a stochastic equation
\begin{equation}
	\partial_t h = \nabla^2 h + (\nabla h)^2 + \eta
	\label{eq:kpz}
\end{equation}
where $\eta$ represents random uncorrelated Gaussian noise.

Equation (\ref{eq:kpz}) has become known as the Kardar-Parisi-Zhang (KPZ) equation \cite{kardar1986dynamic}. It was originally proposed as a continuum model for surface growth due to ballistic deposition \cite{vold1963computer}, since it showed the same dynamic scaling behaviour \cite{family1985scaling}. However, the correspondence between KS and KPZ in two-dimensions has led to disagreements by the same authors  \cite{procaccia1992surface,jayaprakash1993universal,l1994comment,
jayaprakash1994jayaprakash}, because of the lack of conclusive analytical results. More recent results \cite{boghosian1999hydrodynamics,nicoli2010kardar} tend to support the conjecture that KPZ and KS equations belong to the same universality class, although the numerical results for the deterministic (noiseless) KS are not conclusive due to the extremely long transient effects. 
Recent numerical results for the two-dimensional KS \cite{juknevicius2016long} with much longer simulation times show the same scaling properties of the saturated surface roughness as obtained by Manneville and Chat{\'e} \cite{manneville1996phase} for the two-dimentional KPZ, thus, further supporting the argument that the two-dimensional KS and KPZ equations belong to the same universality class.
However, \cite{juknevicius2016long} has found 
different scaling behaviour in the less researched
generalized KS case (\ref{eq:gkse}) with $\alpha>0$.
%%%%%%%%%%%%%%%%%%%%%%%%%%%%%%%%%%

The purpose of this paper is to
demonstrate in a wider parameter range the validity of 
the scaling relations for the surface roughness 
reported in the previous work \cite{juknevicius2016long}, 
and to
investigate the dynamics of the scale-free low-wavenumber spatial variations 
that these relations imply.
%
%%%%%%%%%%%%%%%%%%%%%%%%%%%%%%%%%

The article is structured as follows.
After commenting on numerical methods and parameters used
in the investigation, 
Sec.~\ref{section:roughness} demonstrates 
the initial transient kinetics of the surface roughness 
and eventual cross-over to the saturation regime 
observed in the investigated parameter range.
Sec.~\ref{section:spatial} presents morphologies of
surfaces produced by the generalized KS equation (\ref{eq:gkse}) 
in the saturated regime and
presents results on the finite-size scaling of 
the saturated surface roughness.
The analysis of temporal behaviour of the surface
roughness for the KS case is demonstrated in greater detail
in Sec.~\ref{section:dynamics},
and the results of the same analysis for the generalized-KS case
are then given in Sec.~\ref{section:other}.
We conclude with a summary of our findings and propose
several directions for future work in Sec.~\ref{sec:summary}.

\begin{figure}[ht!]
	\resizebox{0.5\textwidth}{!}{%
	\includegraphics{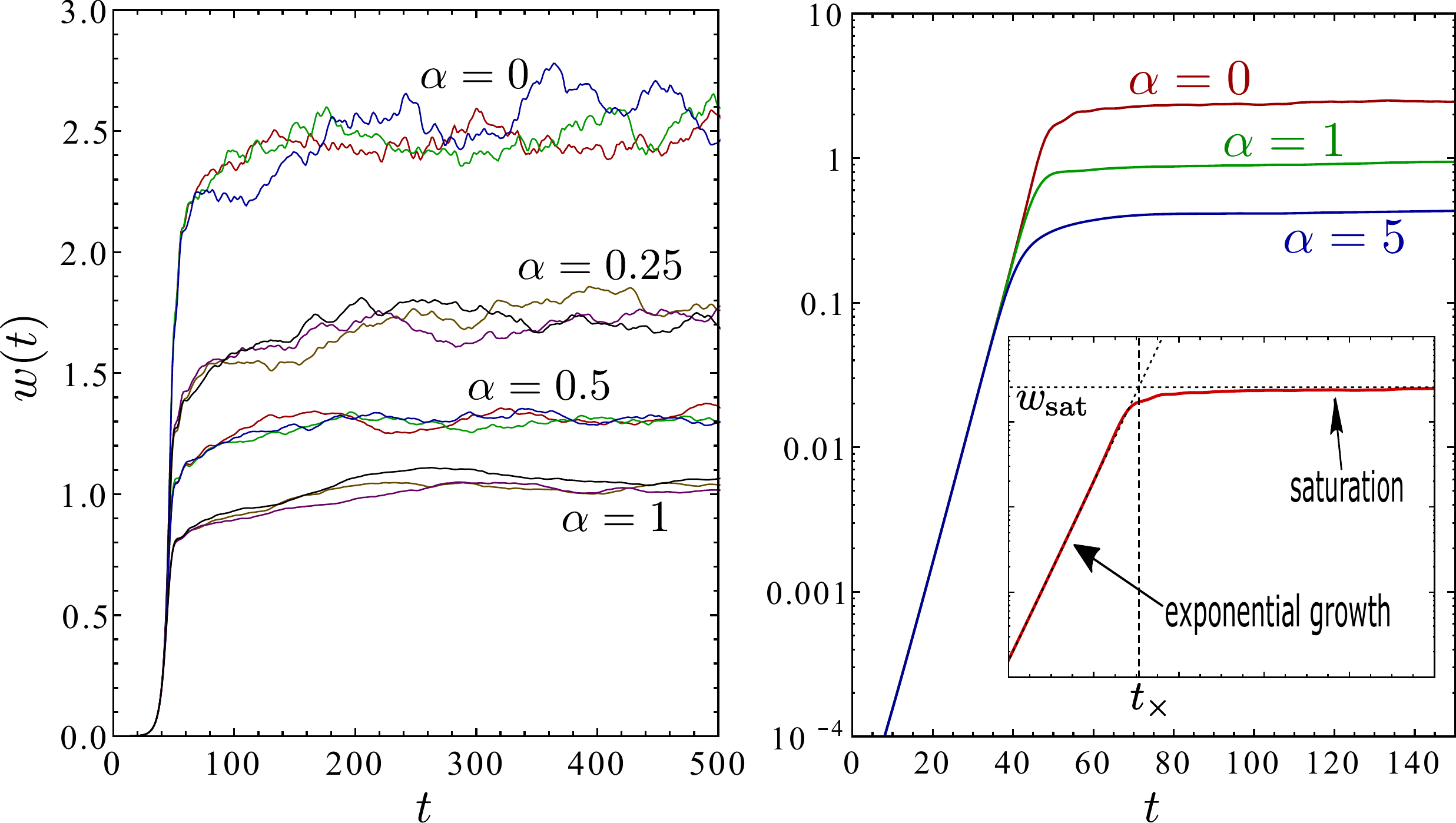}
	}
  \caption{Initial transient kinetics of the roughness $w(t)$, (\ref{eq:rough}), of a surface evolving accourding to (\ref{eq:gkse}) with several values of parameter $\alpha$. Each simulation starts from a random uncorrelated surface profile with initial roughness $w(0)=10^{-4}$. The left panel shows $w(t)$ on linear scale for 3 realizations with each $\alpha$. The right panel highlights the initial exponential increase of $w(t)$ by using 
 a semi-logarithmic scale. The inset demonstrates the saturation roughness $w_{\mathrm{sat}}$ and the cross-over time $t_{\times}$.}
  \label{fig:tievo_anfg}
\end{figure}

\section{Surface roughness: transient kinetics and saturation}
\label{section:roughness}

In this investigation, the equation (\ref{eq:gkse}) is solved numerically 
for different values of 
$\alpha$ using the \emph{finite difference} method with periodic boundary
conditions, the time step $\Delta t=0.005$, and spatial discretization step $\Delta
x =0.71086127010534\,$. 
Such a seemingly bizarre number for the discretization
step $\Delta x$ is actually a good approximation of the value that is needed in
order for the system with periodic boundary conditions to be able to contain
hexagonal patterns that appear in some other versions of the generalized KS equation (see for example \cite{paniconi1997stationary,diddens2015continuum}). The
equation is solved for system sizes $L$ ranging from about 36 to about 711
(i.e., on the $N \times N$ lattices with $N$ from 50 to 1000, where
$L=N\,\Delta x$). Different methods of
numerical solution for (\ref{eq:gkse}) are presented and compared in
\cite{raible2002amorphous}.

In the KS case ($\alpha=0$), the evolving surfaces reach the regime where the dynamics is chaotic, but statistically stationary. This type of behaviour also persists for $\alpha>0$, at least up to $\alpha=5$.
However, for larger values of $\alpha$, this stationary chaotic behaviour gives way to non-stationary effects that prevent the saturation in the surface evolution.
Indeed, in the limiting case when $\alpha\rightarrow\infty$ in (\ref{eq:gkse}), by rescaling $h$, one arrives at the conserved Kuramoto-Sivashinsky equation \cite{raible2000amorphous},
\begin{equation}
	\partial_t h = -\nabla^2 h - \nabla^4 h- \nabla^2(\nabla h)^2\,,
	\label{eq:coarse}
\end{equation}
which produces a non-stationary regime with ever increasing surface roughness due to the uninterrupted coarsening of the surface patterns. Thus, by increasing $\alpha$, there must be a route from the stationary chaotic evolution to non-stationary coarsening behaviour. Nonetheless, the long-time behaviour at the intermediate $\alpha$ values seems to be quite complicated and has not been studied in detail so far.

Even though negative $\alpha$ values do not follow from surface growth or erosion models (see, e.g., \cite{raible2001amorphous,cuerno2011nanoscale}), it
is desirable to understand the dynamics in that range
for completeness.
However, for $\alpha<-0.14$ we find that
large local gradients in the surface emerge and grow.
They eventually exceed the numerical capacity of the simulation, thus making the required long-time calculations unstable.
Therefore, this work focuses on the evolution of surfaces produced by (\ref{eq:gkse}) in a moderate range, $-0.12\leq\alpha\leq5$, of parameter values where the long time behaviour is stationary. 

One of the most important quantities characterizing a surface \cite{tong1994kinetics,barabasi1995fractal} is the surface \emph{roughness} $w(t)$, also called the \emph{surface width}:
\begin{equation}
	w(t):=\sqrt{\Big \langle \big( h(\boldsymbol{r},t)-\bar{h}(t) \big) ^2 \Big\rangle_{\boldsymbol{r}}}\,.
	\label{eq:rough}
\end{equation}
The scaling properties of this quantity are often used to characterize and classify various surface growth models into various universality classes \cite{barabasi1995fractal,family1985scaling,nicoli2010kardar,sneppen1992dynamic}.
The roughness of an evolving surface changes with time. 
In the range of parameter values considered here, the kinetics of
$w(t)$ due to the surface evolution according to (\ref{eq:gkse}) seems
to follow a distinct pattern (see, e.g., \cite{raible2000amorphous, raible2001amorphous,raible2002growth,juknevicius2016long}): starting from a random surface with some small
initial roughness $w(t=0)\ll 1$, the roughness begins to grow at an exponential rate, but at some time $t_{\times}\lesssim 100$ this growth slows down
and, later on, crosses over to a stationary regime where it oscillates about some
average (\emph{saturation}) value $w_{\mathrm{sat}}$. This transient behaviour is shown in Fig.~\ref{fig:tievo_anfg}) for several parameter values.

The value of saturated surface roughness can be defined as follows:
\begin{equation}
	w_{\mathrm{sat}}=\lim_{T\rightarrow\infty}\big\langle w(t) \big\rangle_{t\in[t_{0},\,t_{0}+T)}\,.
	\label{eq:wsat}
\end{equation}
Here $t_0\gg t_{\times}$ is a time at which all initial transient effects have decayed and are virtually undetectable, i.e., the time at which the stationary regime has been reached. In practice, the total observation time $T$ has to be much larger than the typical time scale in the kinetics of $w(t)$.
In the investigation presented here, the saturation values for surface roughness $w_{\mathrm{sat}}$ are calculated using $t_0=2\cdot 10^4$ and $T=8\cdot 10^4$. 
Note that these times are significantly larger than those 
recently achieved by Mu{\~n}oz-Garcia et. al. in 
the numerical investigation of an equation equivalent to (\ref{eq:gkse}) in the one-dimensional case \cite{munoz2006short,munoz2010observation}. There, although the 'interrupted coarsening' is observed, the saturated stationary regime appears not to have been fully reached.

The surface roughness $w(t)$ represents the integral effect of all modes contributing to the surface morphology. 
Therefore, in this work, the time series of chaotic fluctuations of $w(t)$ in the stationary regime are used to investigate the long-time dynamics of surfaces, in particular, the temporal behaviour of the large-scale height variations observed in Ref.~\cite{juknevicius2016long}.
		
\section{Surface morphologies and the scaling of roughness}
\label{section:spatial}

\begin{figure}[ht!]
	\resizebox{0.5\textwidth}{!}{%
	\includegraphics{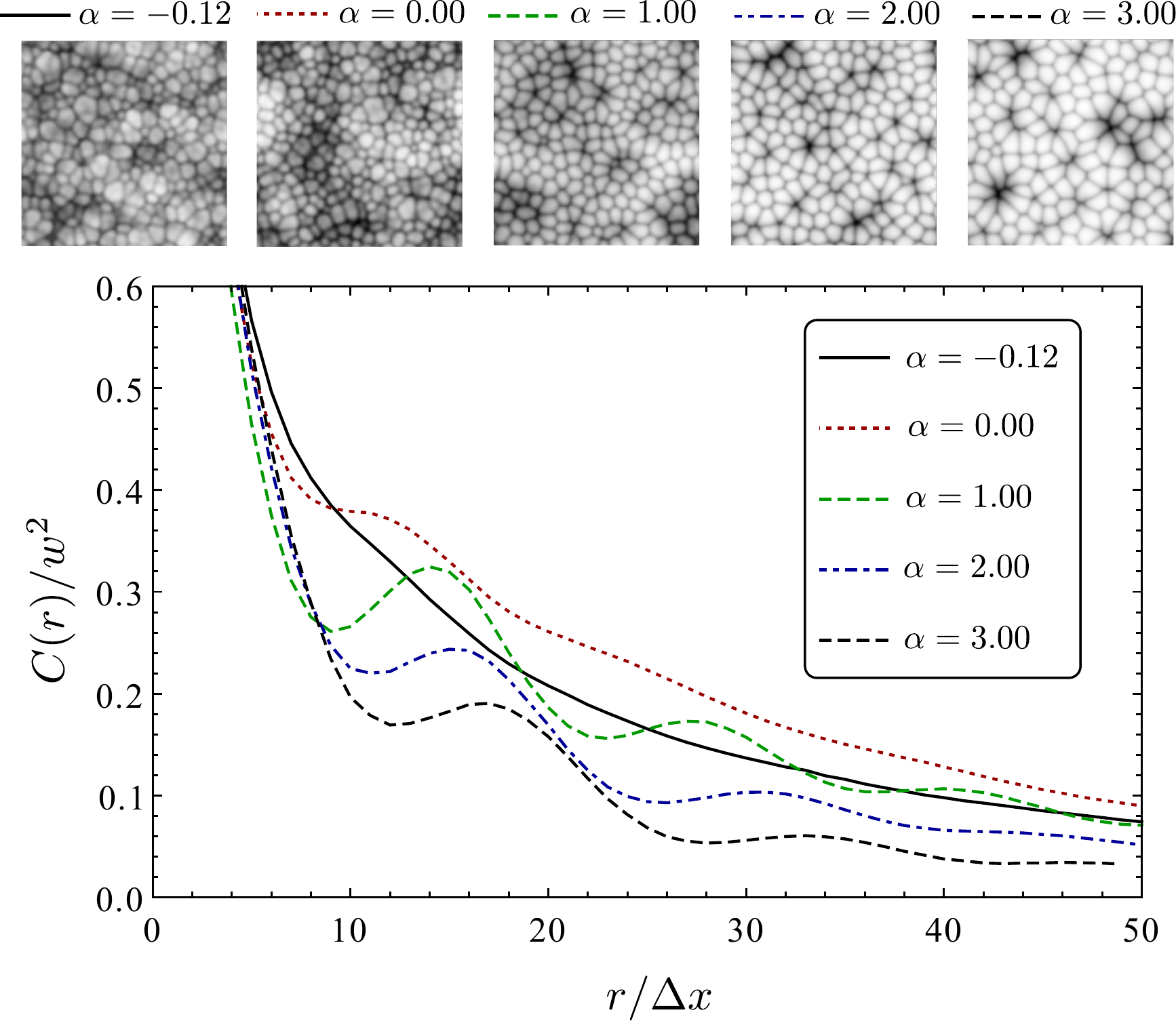}
	}
  \caption{Top panel: Surfaces $h(\boldsymbol{r},t)$ (values of 
  the surface height $h$
    coded in gray-scale) evolving according to (\ref{eq:gkse}) at system size
    $N=200$ ($L\approx142$)  with  parameters $\alpha=-0.12,\,\ldots ,\,3$
    at time $t=10^{5}$ (in the stationary regime). Bottom panel: Normalized
    autocorrelation functions $C(r)$ as defined in (\ref{eq:height-corr-iso}) of
    the surfaces that are shown in the top panel.}
  \label{fig:sucorr_siz300_02}
\end{figure}

\begin{figure}[ht!]
	\resizebox{0.5\textwidth}{!}{%
	\includegraphics{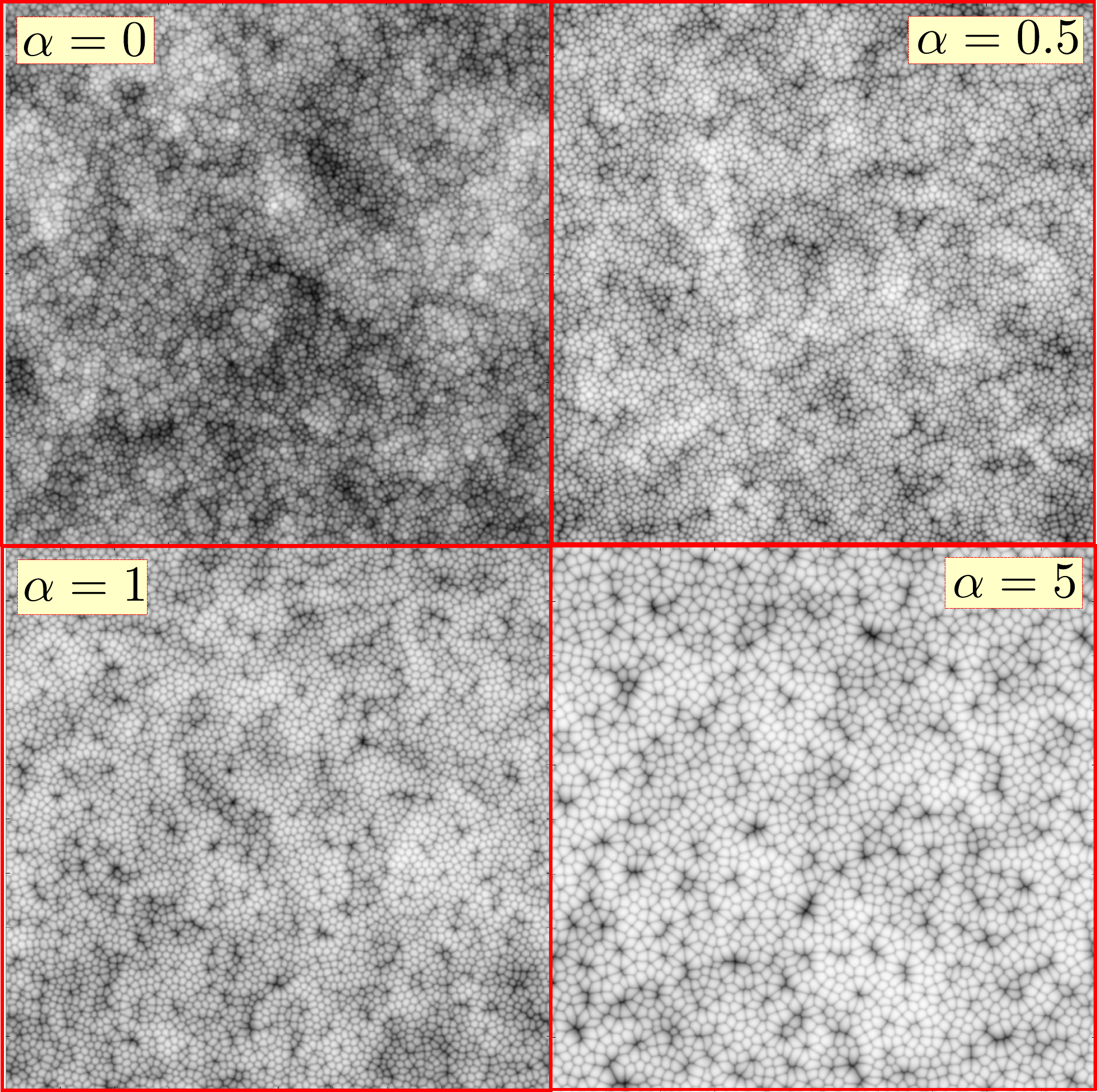}
	}
  \caption{Surfaces (values of the surface height $h$ coded in gray-scale) for the system size $N=1000$
    ($L\approx711$) evolving according to (\ref{eq:gkse}) with  parameters 
    $\alpha=0,\,0.5,\,1,\,5$ at time $t=6 \cdot 10^{4}$.}
	\label{fig:surfaces_siz1000}
\end{figure}

The surface profiles produced by (\ref{eq:gkse}) in the stationary regime have a 
disordered cellular structure \cite{juknevicius2016long, raible2001amorphous} (c.f.~Fig.~\ref{fig:sucorr_siz300_02} and Fig.~\ref{fig:surfaces_siz1000}).
Since (\ref{eq:gkse}) is isotropic, and, consequently, the resulting profiles have no distinct direction on the $\boldsymbol{r}$-plane, the surface morphologies are investigated by averaging the surface
height autocorrelation function over all directions at 
a distance $r=|\boldsymbol{r}|$:
\begin{equation}
	C(r)=\Big\langle \big\langle  (h(\boldsymbol{r'})-\bar{h})\, (h(\boldsymbol{r'}+\boldsymbol{r})-\bar{h}) \big\rangle_{\boldsymbol{r'}} \Big\rangle_{r=|\boldsymbol{r}|}\,.
	\label{eq:height-corr-iso}
\end{equation}
Fig.~\ref{fig:sucorr_siz300_02} shows the resulting surface patterns and the corresponding normalized height correlation
functions $C(r)/w^2$ for relatively small systems (of size
$N=200$, in lattice units) for different parameter
$\alpha$ values. 
The shape of the autocorrelation function at smaller distances gives an insight into the small-scale surface patterns. For example, in Fig.~\ref{fig:sucorr_siz300_02}, one can see how the cellular patterns change, by increasing $\alpha$:
the autocorrelation function (\ref{eq:height-corr-iso}) changes from monotonically decreasing at $\alpha=-0.12$ (corresponding to 'flaky' surface profiles, with 'flakes' of widely varying size) to having a short flat region at $\alpha=0$ (corresponding to a profile with 'cells' of similar size), and to a function with at least one distinct peak at $\alpha>0$ whose distance increases with $\alpha$ (corresponding to the surface 'cells' becoming almost
round 'humps' whose size increases with $\alpha$).

Another thing that can be noticed in Fig.~\ref{fig:sucorr_siz300_02} is that the
normalized correlation function $C(r)/w^2$ decays slowly for $\alpha=0$ and faster for increasing $\alpha$. Also, perhaps surprisingly, the autocorrelation function for $\alpha<0$ decays faster than for $\alpha=0$. These are the first indications of
the influence of parameter $\alpha$ on long-range height correlations.

Simulations show that the resulting saturated surface roughness (\ref{eq:wsat}) increases with the system size. This indicates that the surface profiles of larger systems contain additional spatial Fourier components of smaller wave number $k$, since the structure on smaller scales remains virtually unchanged \cite{juknevicius2016long}.

Large-scale height variations in surfaces produced by (\ref{eq:gkse})
become more distinct as the system size is chosen to be many times larger than
the typical cell size (see Fig.~\ref{fig:surfaces_siz1000}). 

\subsection{Scaling of roughness}
The isotropic surface power-spectral density (PSD), $S(k)$, defined as the absolute square of the Fourier transform of the surface profile integrated over all directions of the wave vectors $\boldsymbol{k}$, can be obtained from the isotropic surface autocorrelation function $C(r)$ \cite{juknevicius2016long,palasantzas1993roughness}:
\begin{equation}
	S(k)=k\,2\pi\, \int\!\mathrm{d}r\,r\,C(r)\, J_0(kr)\,,
	\label{eq:suspec-corr}
\end{equation}
Here $J_0(kr)$ is the Bessel function of the 1st kind:
\begin{equation}
	 J_0(kr)=\frac{1}{2\pi}\int_{0}^{2\pi}\!\mathrm{d}\phi\, \mathrm{e}^{\mathrm{i}kr\cos \phi}\,.
	 \label{eq:bessel}
\end{equation}

\begin{figure}
	{\centering\includegraphics[width=0.45\textwidth]{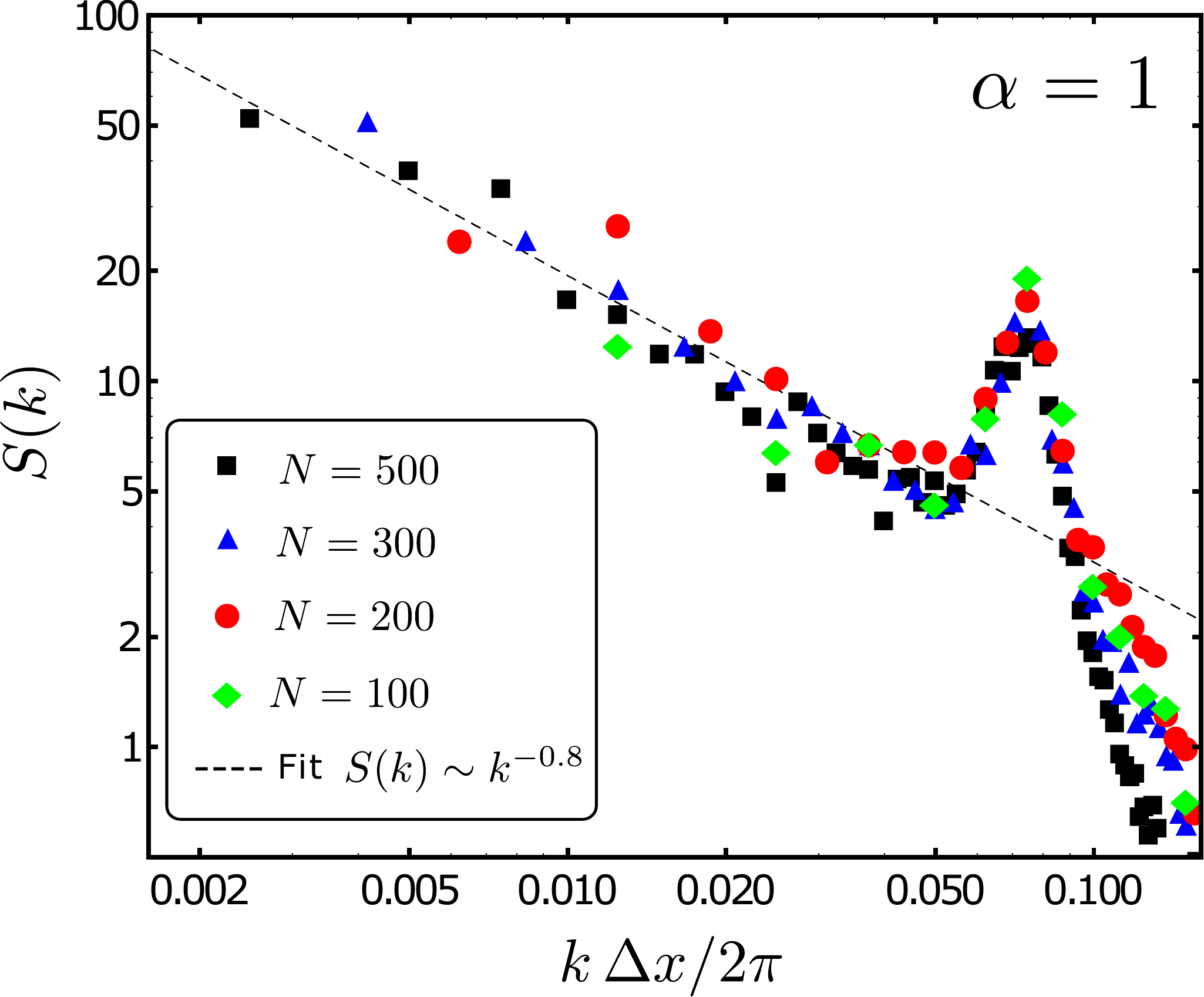}\par}
  \caption{(log-log scale) Numerically calculated PSD of the surfaces at time
    $t=10^5$ produced by (\ref{eq:gkse}) with parameter $\alpha=1$ for system
    sizes (in lattice units) $N=500$ (black squares), $N=300$ (blue triangles), $N=200$ (red circles), and $N=100$ (green diamonds). The black dashed line represents
    power-law fit (\ref{spectrum_pwl}) with exponent $\gamma=0.8$.}
	\label{fig:specsize1}
\end{figure}

An example of numerically calculated surface PSD $S(k)$ using (\ref{eq:height-corr-iso})-(\ref{eq:bessel}) for $\alpha=1$ is shown in Fig.~\ref{fig:specsize1}. Here, one can see a distinct peak that corresponds to an average size of a hump-shaped cells in the surface pattern (see Figs.~\ref{fig:sucorr_siz300_02} and \ref{fig:surfaces_siz1000}) and a power-law trend for small wave numbers.

%\subsection{Finite size effects}

The integral of the PSD $S(k)$ (\ref{eq:suspec-corr}) over all wave numbers $k$
equals the variance of the surface profile which is the square of surface roughness:
\begin{equation}
	\frac{1}{2\pi}\int \!\mathrm{d}k\,S(k)= w^2
	\,\,.
	\label{eq:1d-PSD_int}
\end{equation}
Since the surfaces in numerical simulations are represented on a discrete $(N\times N)$ lattice of finite size $L$ with a discretization step $\Delta x$, wave numbers that can fit into the system are $k_n=n \Delta k$ with $n=1,\,\ldots,\,N$ and $\Delta k = 2\pi/L$.
For large enough systems with $N\gg 1$, according to (\ref{eq:1d-PSD_int}), the square of the surface roughness 
can then be expressed as:
\begin{equation}
	w^2\approx \frac{1}{2\pi}\int_{k_{\mathrm{min}}}^{k_{\mathrm{max}}}\! \!\mathrm{d}k\,S(k)
	\,\,,
	\label{eq:1d-PSD_int-num}
\end{equation}
where
\begin{eqnarray}
	k_{\mathrm{min}}\approx\frac{2\pi}{L}=\frac{2\pi}{N \Delta x}\,,\,\,\,\,\nonumber&
	\label{eq:minmax_wavenumber}
	k_{\mathrm{max}}\approx\frac{2\pi}{ \Delta x}\,.\nonumber
\end{eqnarray}

If the discretization step $\Delta x$ is kept constant (implying $k_{\mathrm{max}}=\mathrm{const}$), and the surface
patterns at different system sizes $L$ (up to the smallest wave number $k_{\mathrm{min}}\propto L^{-1}$) remain statistically the same (as in Fig.~\ref{fig:specsize1}), then, by increasing the system size $L$, the calculated dependence $w^2(L)$ should yield, according to (\ref{eq:minmax_wavenumber}), the shape of the surface PSD $S(k)$ for small wave numbers $k\rightarrow 0$. 
This is useful, since, for larger systems, the direct calculation of the two-dimensional autocorrelation function (\ref{eq:height-corr-iso}) and surface spectrum
(\ref{eq:suspec-corr}) can take a very long computation time.

%\subsection{Power-law spectrum at small wave numbers}
In \cite{juknevicius2016long}, an assumption was made that the PSD $S(k)$ (\ref{eq:suspec-corr}) of surfaces produced by (\ref{eq:gkse}) has a power-law shape for small wave numbers (below some value $k_{\mathrm{s}}$):
\begin{equation}
	S(k)=
		C\,k^{-\gamma} \,\,\, \mathrm{for} \,\,\, k < k_{\mathrm{s}}\,.
	\label{spectrum_pwl}
\end{equation}

By substituting (\ref{spectrum_pwl}) into (\ref{eq:1d-PSD_int-num}), one gets three qualitatively distinct scaling behaviours $w^2(L)$ for 
$L>2\pi\,k_{\mathrm{s}}^{-1}$, 
depending on the value of spectral exponent $\gamma$ in (\ref{spectrum_pwl}):

\begin{equation}
		\left\{
		\begin{array}{lcc}
			w^2(L) = C_1 - C_2\,L^{-(1-\gamma)} & \mathrm{for} & \gamma<1\\
			w^2(L)= C \ln L + B & \mathrm{for} & \gamma=1\\
			w^2(L)= D_1 + D_2\,L^{\gamma-1} & \mathrm{for} & \gamma>1
		\end{array}
		\right.
		\label{wsquare_scaling}
\end{equation}	

For asymptotically large systems $L\rightarrow \infty$, 
(\ref{wsquare_scaling}) would become 
\begin{equation}
		\left\{
		\begin{array}{lcc}
			w^2 (L) \sim \mathrm{const} & \mathrm{for} & \gamma<1\\
			w^2 (L) \sim \ln L & \mathrm{for} & \gamma=1\\
			w^2 (L)\sim L^{\gamma-1} & \mathrm{for} & \gamma>1\,,
		\end{array}
		\right.
		\label{wsquare_asympt}
\end{equation}	
corresponding to asymptotically constant roughness for $\gamma<1$, logarithmically increasing square of the surface roughness for $\gamma=1$, and power-law scaling for $\gamma>1$.

It has been shown in \cite{juknevicius2016long} that the assumption (\ref{spectrum_pwl}) of a power-law surface PSD with ($0<\gamma\leq 1$) at small wave numbers is indeed valid for surfaces produced by (\ref{eq:gkse}) with parameter values $0\leq\alpha\leq 1$, since the relations (\ref{wsquare_scaling}) fit the numerically calculated surface roughness exceptionally well. 

\begin{figure}[ht!]
  {\centering\includegraphics[width=0.45\textwidth]{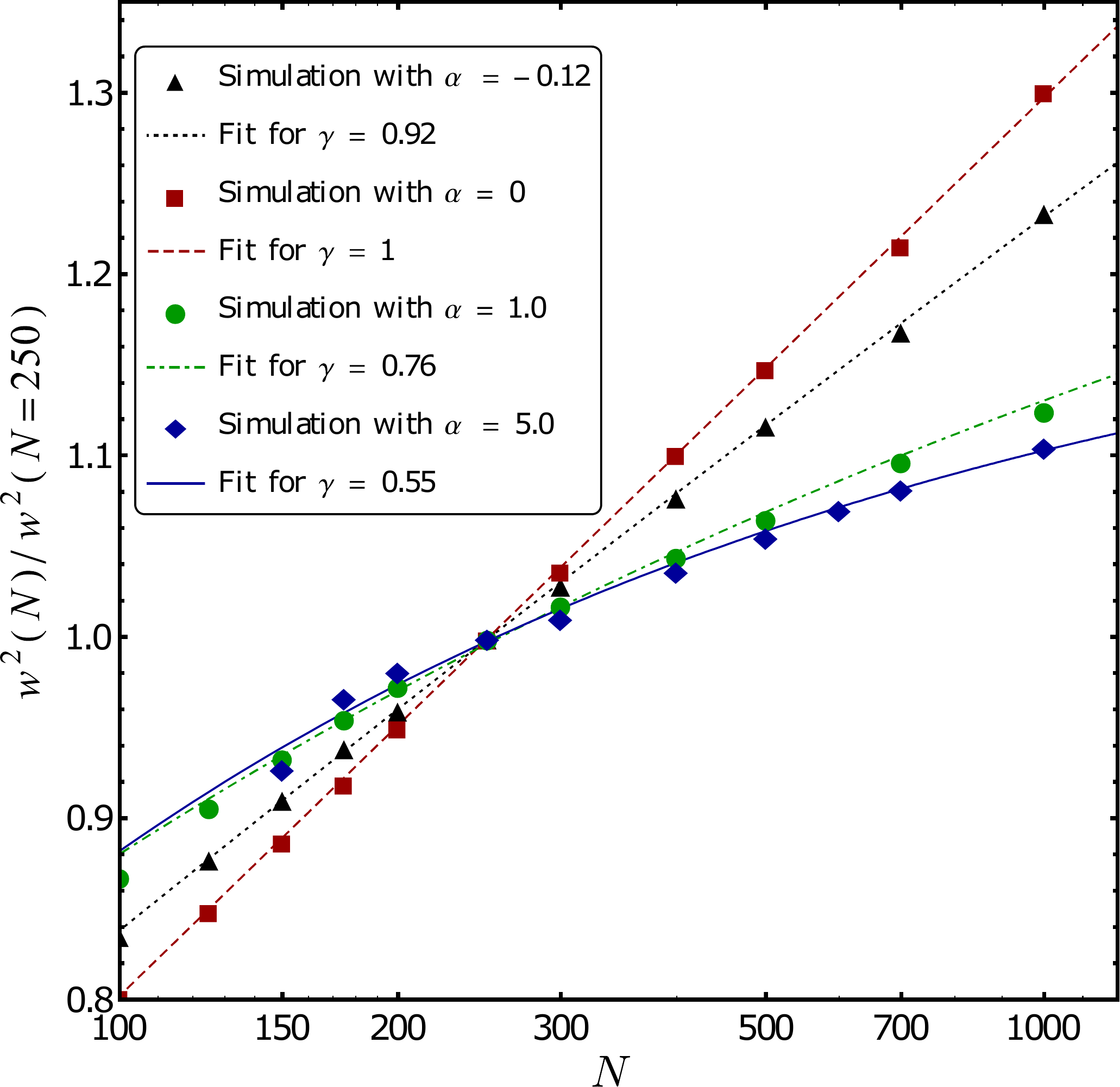}\par}
  \caption{(log-linear scale) Time averaged square of the normalized surface roughness $w^2$ plotted as a function of the system size $N$ (in lattice units). 
Symbols: numerical results for surfaces evolving according to (\ref{eq:gkse}) with different $\alpha$ values.
Lines: fits of the numerical results by (\ref{wsquare_scaling}).}
  \label{fig:scaling01}
\end{figure}

Investigations of a broader parameter range, $-0.12\leq\alpha\leq 5$, presented in this paper, show that the same assumption (\ref{spectrum_pwl}) also holds for other parameter values. 
Fig.~\ref{fig:scaling01} shows the calculated square of the surface roughness $w^2$ dependence on the system size $L=N \Delta x$.
In order to fit the results with different $\alpha$ values in the same plot, the numerical results and their fits for each $\alpha$ have been divided by the corresponding $w^2$ values at $N=250$. 
At $\alpha=0$ the resulting spectral exponent $\gamma=1$ gives the logarithmic dependence $w^2(N)$ (see (\ref{wsquare_scaling})) which is a straight line in 
the log-linear scale.
%(logarithmic scale for abscissa and linear scale for ordinate).
This scaling is the same as found by Manneville and Chat{\'e} for the two-dimensional KPZ equation \cite{manneville1996phase}.

As the parameter increases from $\alpha=0$ to $\alpha=5$, the $\gamma$ values are found to decrease from $\gamma=1$ to $\gamma\approx 0.55$ (see Fig.~\ref{fig:scaling01}). This corresponds to slower-than-linear growth of $w^2$ with $\ln N$.
Hence, for large systems $w^2$ approaches a finite
value.
Perhaps unexpectedly, for $\alpha<0$, the exponent $\gamma$
has also been observed to become smaller than $1$.
Therefore, we conclude that the scaling properties of 
the generalized KS equation 
(\ref{eq:gkse}) differs from those of the KPZ equation
when $\alpha\neq 0$.

\section{Analysis of roughness dynamics in the Kuramoto-Sivashinsky case}
\label{section:dynamics}

Model equation (\ref{eq:gkse}) produces disordered spatial patterns that evolve in time.
As shown in Sec.~\ref{section:spatial}, 
with an increase of the system size $L$
new long range height variations appear in the resulting surface profiles in addition to the small scale patterns.
The apparent scale-free character of these slow height variations is very different from the cellular patterns on small scales which have a characteristic length (the average size of a 'cell' or 'hump'). Also, the spatial properties of both, the small scale patterns and the large scale height variations, depend strongly on the value of parameter $\alpha$ in (\ref{eq:gkse}). This section investigates the corresponding dynamics of these surfaces.

In order to understand the complex spatio-temporal behaviour of (\ref{eq:gkse}), we investigate the dynamics of surfaces it produces by analysing the numerically obtained time series of the surface roughness $w(t)$ which contains the collective behaviour of all modes.
The time series of $w(t)$ are investigated in the time interval $t\in [2\cdot10^4,10^5)$ with sampling time $\tau_{\mathrm{sample}}=1$ (i.e., sampled every $200$ time steps $\Delta t=0.005$), that is, $8\cdot10^4$ values in total for every realization. The results are averaged over 5 to 10 realizations (differing in the initial surface profile) for every parameter $\alpha$ value. For the range of parameter values explored here, the surface evolution can be considered stationary and ergodic, since the statistical properties of $w(t)$ (average, standard deviation, skewness, autocorrelation function) seem to vary little from realization to realization. Moreover, their values calculated in large enough subintervals of the total time interval differ only slightly from each other.

In this section, the analysis of $w(t)$ is presented in more detail for parameter value $\alpha=0$, that is, the Kuramoto-Sivashinsky case (\ref{eq:kse}). The same analysis performed on other parameter values is discussed in Sec.~\ref{section:other}.

\begin{figure}[ht!]
  {\centering\includegraphics[width=0.5\textwidth]{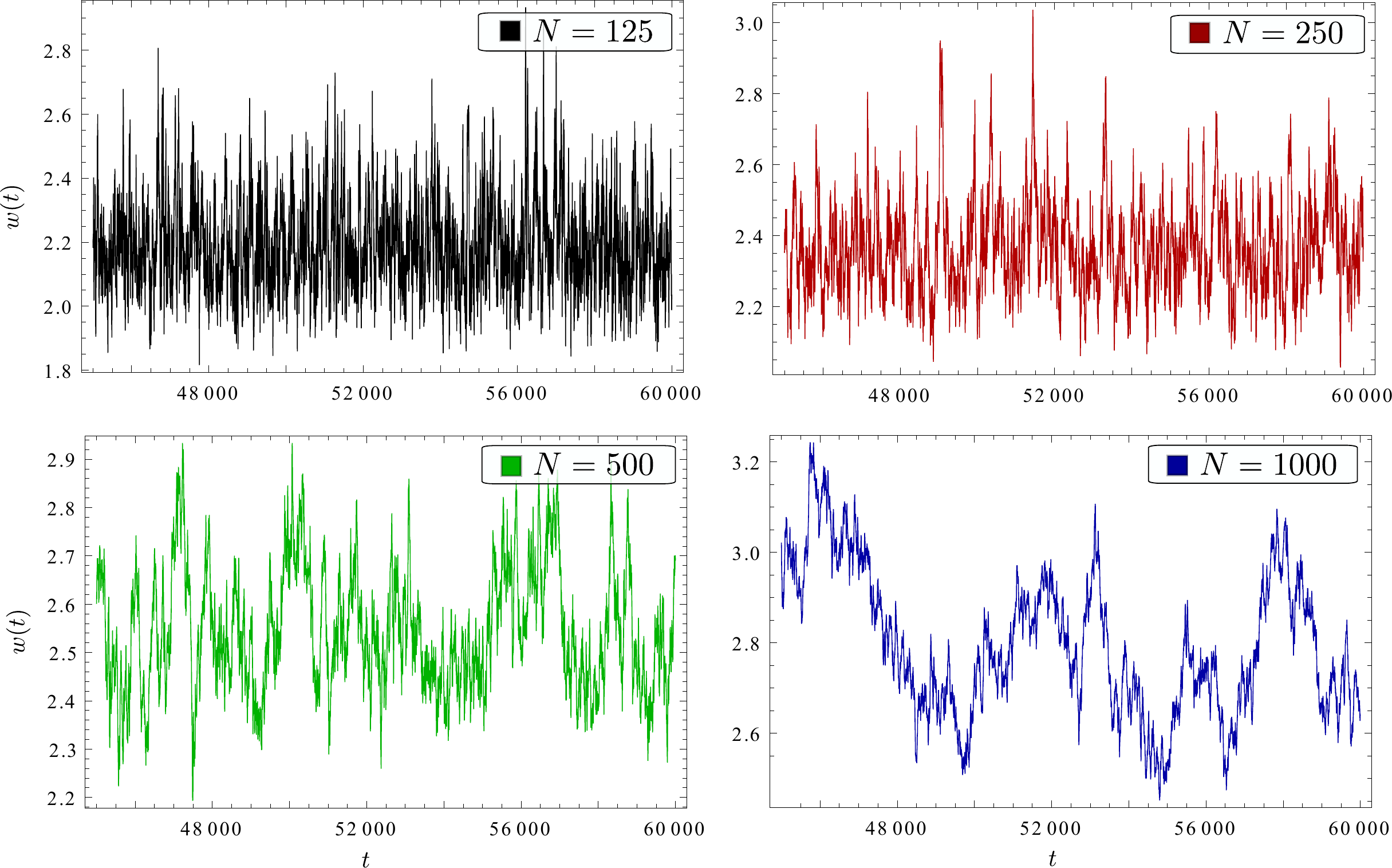}\par}
  \caption{Time series of the surface roughness $w(t)$, $t\in [4.5\cdot10^4,6\cdot10^4)$ for $\alpha=0$ and different system sizes $N$ (in lattice units).}
  \label{fig:tievo000sizes}
\end{figure}

\subsection{Occurrence of slow modes}
Fig.~\ref{fig:tievo000sizes} shows a representative sample
of a surface roughness $w(t)$ time series for $\alpha=0$ and system sizes varying from $N=125$ to $N=1000$. 
Even though the roughness dynamics is dominated
by white noise for small systems ($N=125$), additional slow modes appear as the system size is increased. For relatively large systems ($N=1000$), the
time series in question is similar to a signal 
produced by a random walk. 

\begin{figure}[ht!]
  {\centering\includegraphics[width=0.5\textwidth]{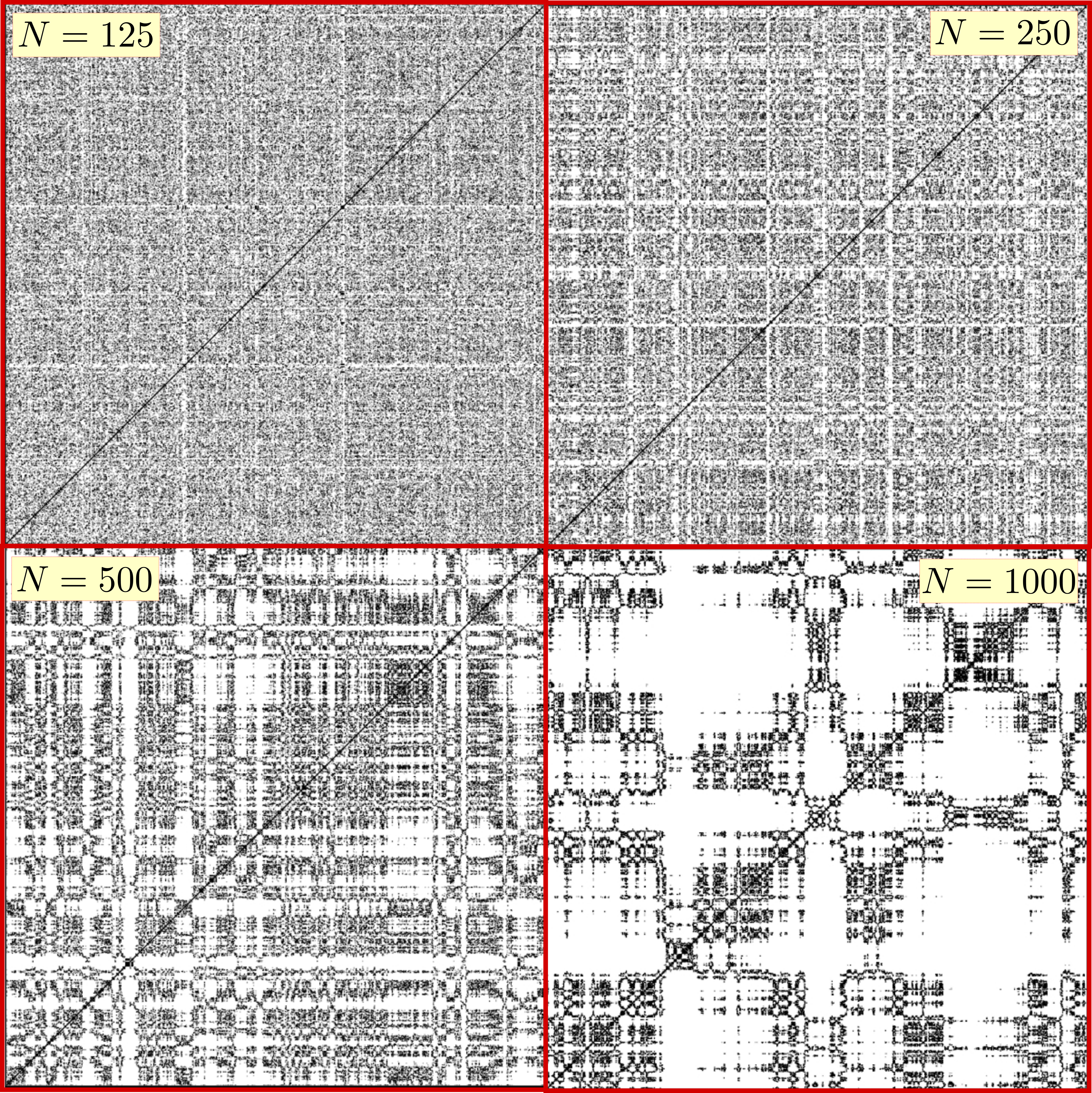}\par}
  \caption{Recurrence plots of the one-dimensional time series 
  of the surface roughness $w(t)$ for $t\in [8\cdot10^4,10^5)$,
  $\alpha=0$, and different system sizes $N$ (in lattice units). }
  \label{fig:recplots000sizes}
\end{figure}

This transition can be visualized even more clearly by using the recurrence plot technique 
\cite{eckmann1987recurrence,gao2000structures,marwan2007recurrence,kantz2004nonlinear} 
(see Fig.~\ref{fig:recplots000sizes}). 
There, a time series $s(t)$ is depicted by plotting a matrix $\boldsymbol{R}_{t_i,t_j}$. In the plot, the 
axes represent the discrete time $t_i$ and $t_j$. A black dot ($\boldsymbol{R}_{t_i,t_j}=1$) is put at a point $(t_i,t_j)$ 
if the values of the time series $s(t)$ at these times 
coincide (recur) to a given accuracy 
$\epsilon$. The pixel
remains white otherwise (value $\boldsymbol{R}_{t_i,t_j}=0$), that is:
 \begin{equation}
 	\boldsymbol{R}_{t_i,t_j}=\Theta(\epsilon-|s(t_i)-s(t_j)|)\,,
 	\label{eq:recplot1}
 \end{equation}
where $\Theta(x)$ is the Heaviside step function.
Each of the recurrence plots in Fig.~\ref{fig:recplots000sizes} is made for a single realization of $w(t)$ in the time interval $t\in [8\cdot10^4,10^5)$, i.e., one fourth 
of the total length of the time series is investigated.

The slow fluctuations of $w(t)$ that appear when the system size is increased can be attributed to the low wave number spatial modes that occur in larger systems. By investigating the scaling properties of these fluctuations, connections between spatial and temporal properties of the corresponding large scale height variations can be made.

\subsection{Autocorrelation functions}
 %The time series (TS) of the surface roughness $w(t)$ resulting from 
 %(\ref{eq:gkse}) in the stationary regime show different properties for different %system sizes $L$ (or $N$, in lattice units), as is clear from Figs.~\ref{fig:tievo000sizes} and \ref{fig:recplots000sizes}. In particular, slow fluctuations appear in $w(t)$ for large systems.

\begin{figure}[ht!]
  {\centering\includegraphics[width=0.45\textwidth]{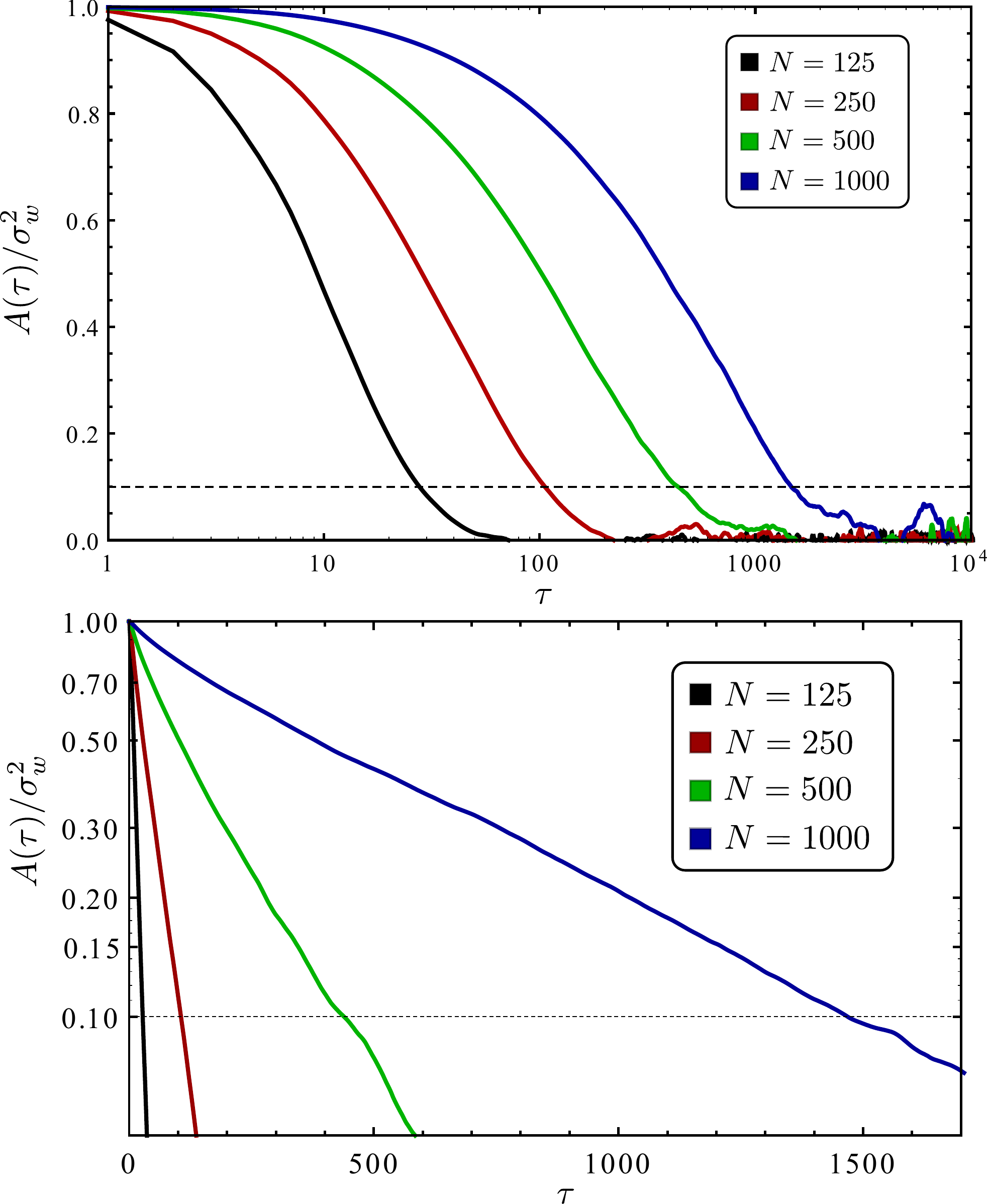}\par}
  \caption{Normalized autocorrelation functions $A(\tau)$ of 
  the surface roughness $w(t)$, $t\in [2\cdot10^4,10^5)$ for $\alpha=0$ and different system sizes (in lattice units) $N$. 
  Top panel: log-linear scale. Bottom panel: semi-logarithmic scale. Horizontal dashed and dotted lines in both panels represent the $A(\tau)/\sigma^2_w=0.1$. }
  \label{fig:tiacorr000sizes}
\end{figure}

The character of the slow fluctuations that appear in the time 
series (TS) of the surface roughness $w(t)$ resulting from 
(\ref{eq:gkse}) for large systems (see Figs.~\ref{fig:tievo000sizes} 
and \ref{fig:recplots000sizes}) is captured by
their autocorrelation functions,
\begin{equation}
	A(\tau)=\big\langle (w(t)-\bar{w})(w(t+\tau)-\bar{w}) \big\rangle_{t}\,,
	\label{eq:tiacorr_def}
\end{equation}
where $\bar{w}=\langle w(t)\rangle_t$ is the average value of $w(t)$ in the stationary regime. Fig.~\ref{fig:tiacorr000sizes} shows the autocorrelation functions obtained from TS of $w(t)$ with $\alpha=0$ for four different system sizes increasing by the factor of $2$: $N=125,\,250,\,500,\,1000\,$.
In the top panel of Fig.~\ref{fig:tiacorr000sizes}, the normalized (i.e., divided by the variance $\sigma_{w}^2=\langle (w(t)-\bar{w})^2\rangle_t\equiv A(0)$) autocorrelation functions are displayed in the log-linear scale.
%where the axis of the lag time $\tau$ (abscissa) is logarithmic, 
%and the axis of $A(\tau)/\sigma_{w}^2$ (ordinate) is linear. 
In this plot, one can immediately recognize the way in which the 
characteristic time scales in $w(t)$ grow with $N$.
For instance, by defining some characteristic correlation time $\tau_{\mathrm{corr}}$ as, for example, the lag $\tau$ at which the autocorrelation function decays to the $10\%$ 
(dashed horizontal line) of its initial value at $\tau=0$, i.e.,
\begin{equation}
	\tau_{\mathrm{corr}}=\min \{\tau>0\, |\, A(\tau)/\sigma_{w}^2\leq 0.1\}\,,
	\label{eq:corrtime_def}
\end{equation} 
one can see that it increases by about the same factor (corresponding to almost constant shifts along a logarithmic scale of $\tau$ axis) as the system size $N$ increases by a factor of $2$. This indicates that the characteristic time $\tau_{\mathrm{corr}}$ grows as a power law of $N$:
\begin{equation}
	\tau_{\mathrm{corr}}\propto N^{\xi}\,.
	\label{eq:corrtime_pwl}
\end{equation} 

Further insight into the dynamics can be gained by looking at
the same autocorrelation functions in a semi-logarithmic
plot, as displayed on the bottom panel of Fig.~
\ref{fig:tiacorr000sizes}. 
%Here, the functions are plotted with $
%\tau$-axis(abscissa) linear and $A(\tau)/\sigma_{w}^2$-axis 
%(ordinate) logarithmic. 
Plotted this way, the autocorrelation 
functions $A(\tau)$ appear almost as straight lines (with an 
additional kink at very small $\tau$) indicating that the their 
shape should be approximately exponential:
\begin{equation}
	A(\tau)\approx \sigma^2_w \, \mathrm{e}^{-\lambda |\tau|}\,.
	\label{eq:tiacorr_exp}
\end{equation}

\subsection{Power spectra and characteristic frequencies}

In order to obtain more quantitative results, it is essential to look at the shape of the corresponding power spectra of $w(t)$.
As stated by the Wiener-Khinchin theorem \cite{yaglom2004introduction}, the power spectral density (PSD) $W(f)$ of a signal can be obtained by Fourier transforming its autocorrelation function (\ref{eq:tiacorr_def}):
\begin{equation}
	W(f)=
	\int_{-\infty}^{\infty}\!\!\mathrm{d}\tau\,A(\tau)\,
	\mathrm{e}^{-\mathrm{i}2\pi f \tau}\,.
	\label{eq:tiacorr_ft}
\end{equation}
By substituting the exponentially decaying autocorrelation function $A(\tau)\propto\mathrm{e}^{-\lambda |\tau|}$ (as in (\ref{eq:tiacorr_exp})) into (\ref{eq:tiacorr_ft}), the PSD $W(f)$ of a Lorentzian shape is obtained:
\begin{equation}
	W(f)\propto \frac{f_0}{f_0^2 + f^2}\,,
	\label{eq:lorentzian}
\end{equation}
where $f_0=\lambda / 2\pi$ is the characteristic frequency that signifies the cross-over between different behaviours of $W(f)$, namely:
\begin{equation}
	W(f)\sim 
	\left\{
	\begin{array}{cll}
		 \mathrm{const} &  ,\,\,  & f \ll f_0\\
		 f^{-2}         &  ,\,\,  & f \gg f_0 \,\,.
	\end{array}
	\right.
\end{equation}
Thus, $f_0$ represents the lowest frequency (or the lowest decay rate $\lambda\propto f_0$) that affects the dynamics of $w(t)$. The above considerations suggest that $f_0$ must correspond to the lowest wave number, $k_{\mathrm{min}}\propto L^{-1}$, 
of a spatial mode occurring in the system of size $L$. 

\begin{figure}[ht!]
  {\centering\includegraphics[width=0.45\textwidth]{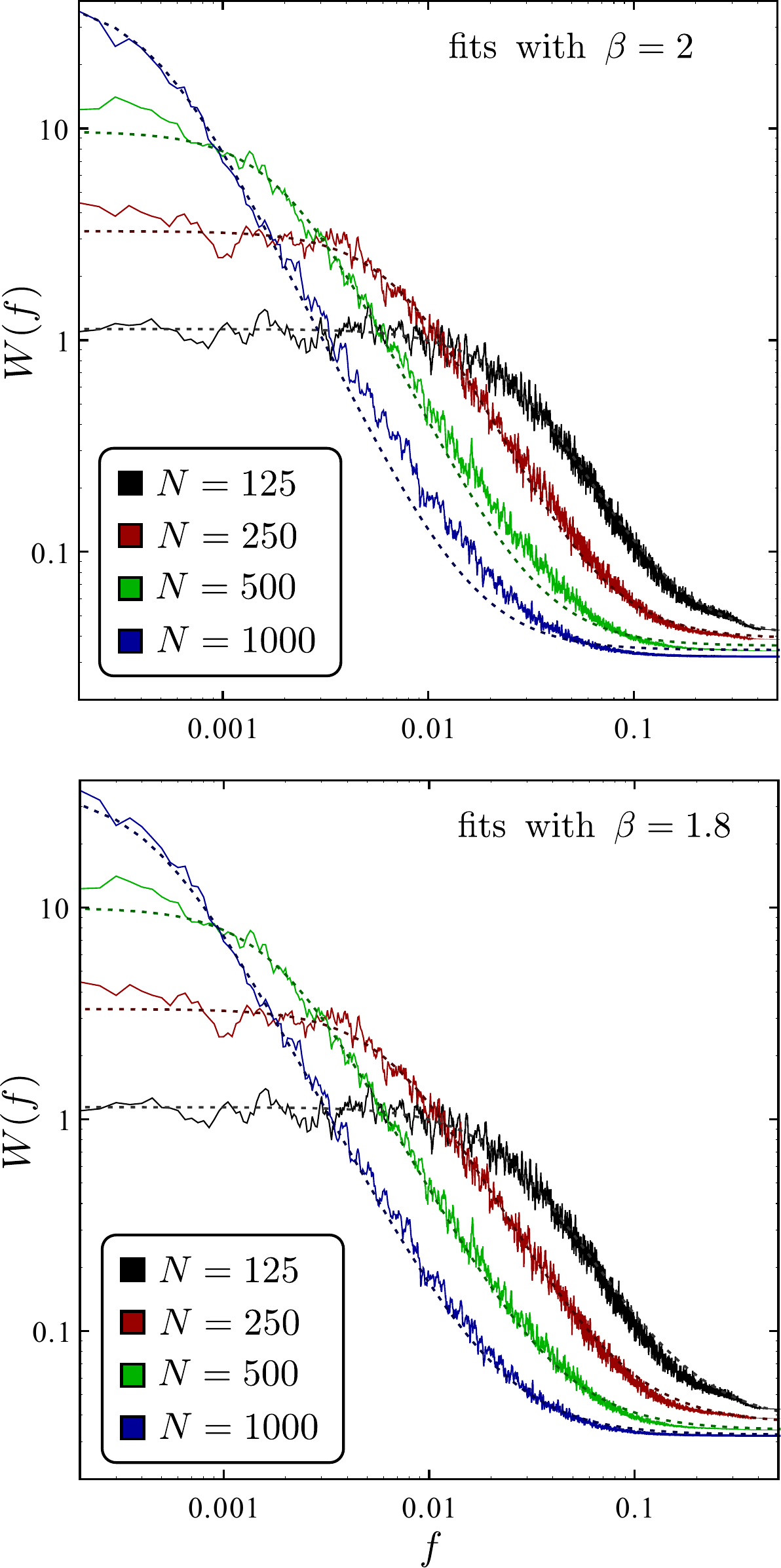}\par}
  \caption{The power spectral densities $W(f)$ of 
  the surface roughness $w(t)$ for $t\in [2\cdot10^4,10^5)$, $\alpha=0$, and different system sizes $N$. Solid lines: calculation results of (\ref{eq:tiacorr_ft}) with (\ref{eq:tiacorr_def}). Dotted lines: fits of the calculated results by (\ref{eq:gen-lorentzian_fit}) with $\beta=2$ (top panel) and $\beta=1.8$ (bottom panel).}
  \label{fig:tispec000_fits}
\end{figure}

Plotted in semi-logarithmic scale (bottom panel of Fig.~\ref{fig:tiacorr000sizes}), the autocorrelation functions $A(\tau)$ appear as almost straight lines corresponding to the approximate exponential decay (\ref{eq:tiacorr_exp}) whose PSD is a Lorentzian (\ref{eq:lorentzian}). Nevertheless, there are deviations from this trend at very short lag times $\tau$. These deviations correspond to additional fluctuations with a very short correlation time -- a \emph{white noise} whose PSD is a constant. 
Therefore, the resulting PSD of $w(t)$ can be fitted by a Lorentzian plus a constant:
\begin{equation}
	W_{\mathrm{fit}}(f)=\frac{A}{f_0^2 + f^2} + B\,
	\label{eq:lorentzian_fit}
\end{equation}
where $A$, $B$ and $f_0$ are fit parameters. 

The PSDs obtained from the autocorrelation functions 
of the surface roughness at $\alpha=0$ for different system sizes $N=L/\Delta x$ are shown in the top panel of Fig.~\ref{fig:tispec000_fits} together with their fits by (\ref{eq:lorentzian_fit}).
A closer analysis shows that a function with a generalized Lorentzian plus a constant $B$,
\begin{equation}
	W_{\mathrm{fit}}(f)=\frac{A}{(f_0^2 + f^2)^{\beta/2}} + B\,,
	\label{eq:gen-lorentzian_fit}
\end{equation}
with $\beta=1.8$ fits the calculated PSDs even better (see the bottom panel of Fig.~\ref{fig:tispec000_fits}).

\begin{figure}[ht!]
  {\centering\includegraphics[width=0.45\textwidth]{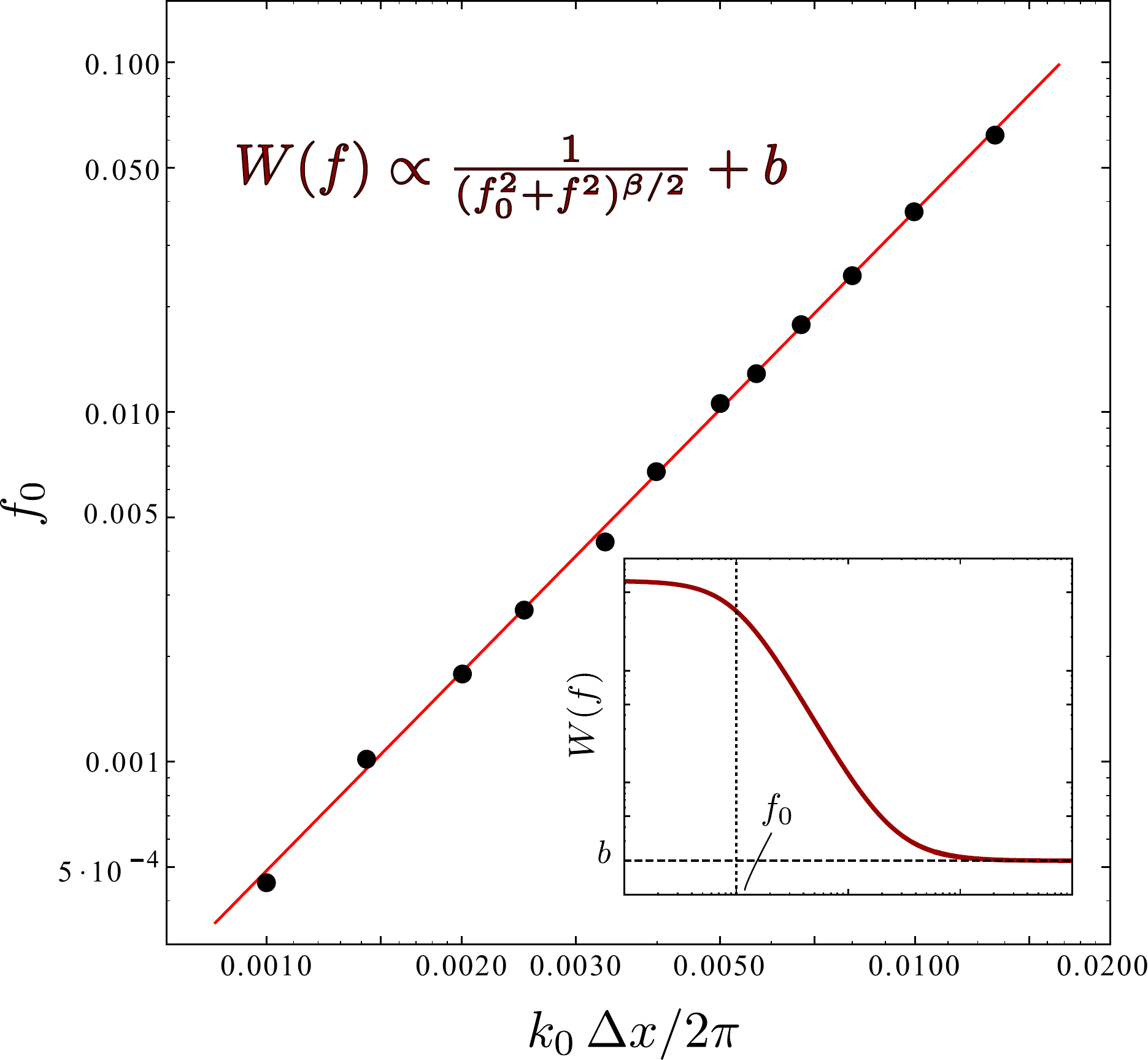}\par}
  \caption{(log-log scale) Relation between the lowest frequency $f_0$ and the lowest wave number $k_0$ occurring in the system for $\alpha=0$. The black filled circles are $f_0$ values obtained as fit parameters for numerical results by (\ref{eq:gen-lorentzian_fit}). The solid red line is the power-law fit $f_0 \propto k_0^{\xi}$ with $\xi\approx 1.89$. The inset on the bottom right shows $f_0$ for the PSD $W(f)$ defined by the expression shown on the top left.}
  \label{fig:dispers000}
\end{figure}

The cross-over frequency $f_0$ obtained as a fit parameter represents the lowest frequency (corresponding to the longest time scale) in the kinetics of $w(t)$. In Fig.~\ref{fig:tispec000_fits}, it is clearly visible that $f_0$ decreases as the system size $N$ is increased. Since the lowest wave number $k_{\mathrm{min}}$ of the spatial modes occurring in the system is inversely proportional to the system size, $k_{\mathrm{min}}\propto N^{-1}$, the $f_0(N)$ dependence connects the spatial and the temporal scales.
Indeed, by defining the some critical wave number $k_0$ as
\begin{equation}
	k_0=\frac{2\pi}{L}\equiv\frac{2\pi}{\Delta x}\, \frac{1}{N}\propto k_{\mathrm{min}}\,,
\end{equation}
one can obtain a dispersion relation $f_0(k_0)$ --- a connection between the lowest wave number in the system and its corresponding frequency. The resulting $f_0$ dependence on $k_0 \Delta x /(2\pi)=N^{-1}$ for $\alpha=0$ is shown in Fig.~\ref{fig:dispers000} in the double-logarithmic scale. Plotted this way, the results appear to lie on a straight line, meaning that the relation is approximately a power-law
$
	f_0\propto k_0 ^ {\,\,\xi}\,
$
with the exponent $\xi\approx 1.89$, as the fit shows
(c.f.~Fig.~\ref{fig:dispers000}).

\section{Dynamics of roughness for other parameter values}
\label{section:other}

\begin{figure}[ht!]
  {\centering\includegraphics[width=0.5\textwidth]{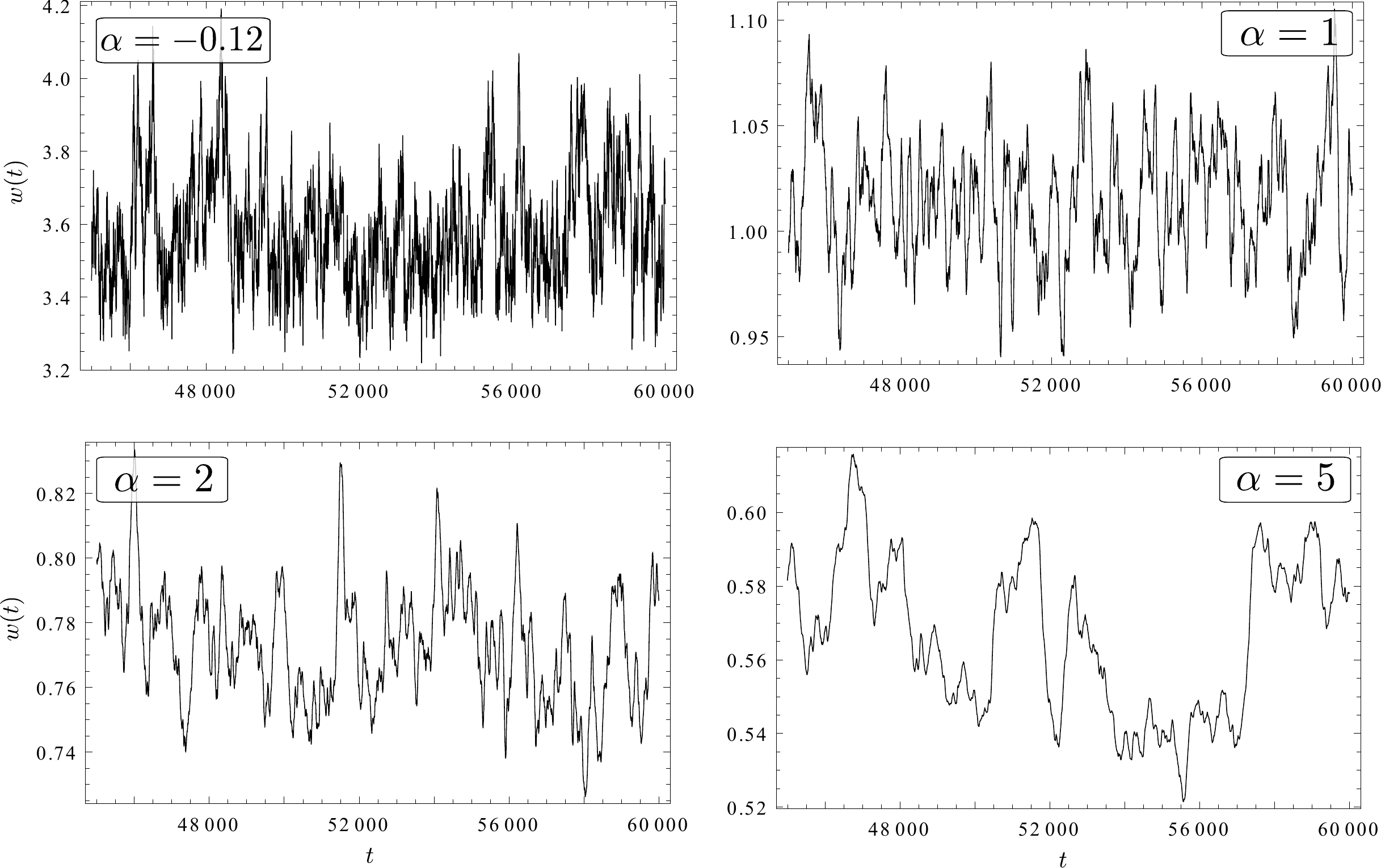}\par}
  \caption{Time series of the surface roughness $w(t)$
  for $t\in [4.5\cdot10^4,6\cdot10^4)$, system size $N=500$ (in lattice units), and different values of parameter $\alpha$.}
  \label{fig:tievo_siz500alphas}
\end{figure}

The fluctuations of $w(t)$ change character as parameter $\alpha$ is varied. This can already be seen from their time series (Fig.~\ref{fig:tievo_siz500alphas}). This section presents some of the results on spatio-temporal properties of surfaces evolving according to (\ref{eq:gkse}) with parameter values $\alpha\neq 0$ in order to point out the similarities and differences from the $\alpha=0$ case presented in Sec.~\ref{section:dynamics}.

\subsection{Spatio-temporal properties}
The same type of analysis, as presented in Sec.~\ref{section:dynamics} for parameter $\alpha=0$, has also been performed for other parameter values.

As in the $\alpha=0$ case, for $\alpha\neq 0$, the occurrence of slow modes can also be observed as the system size increases. However, since the character of low wave number spatial variations depends on $\alpha$, as shown in Sec.~\ref{section:spatial}, their temporal properties also differ.

\begin{figure}[ht!]
  {\centering\includegraphics[width=0.5\textwidth]{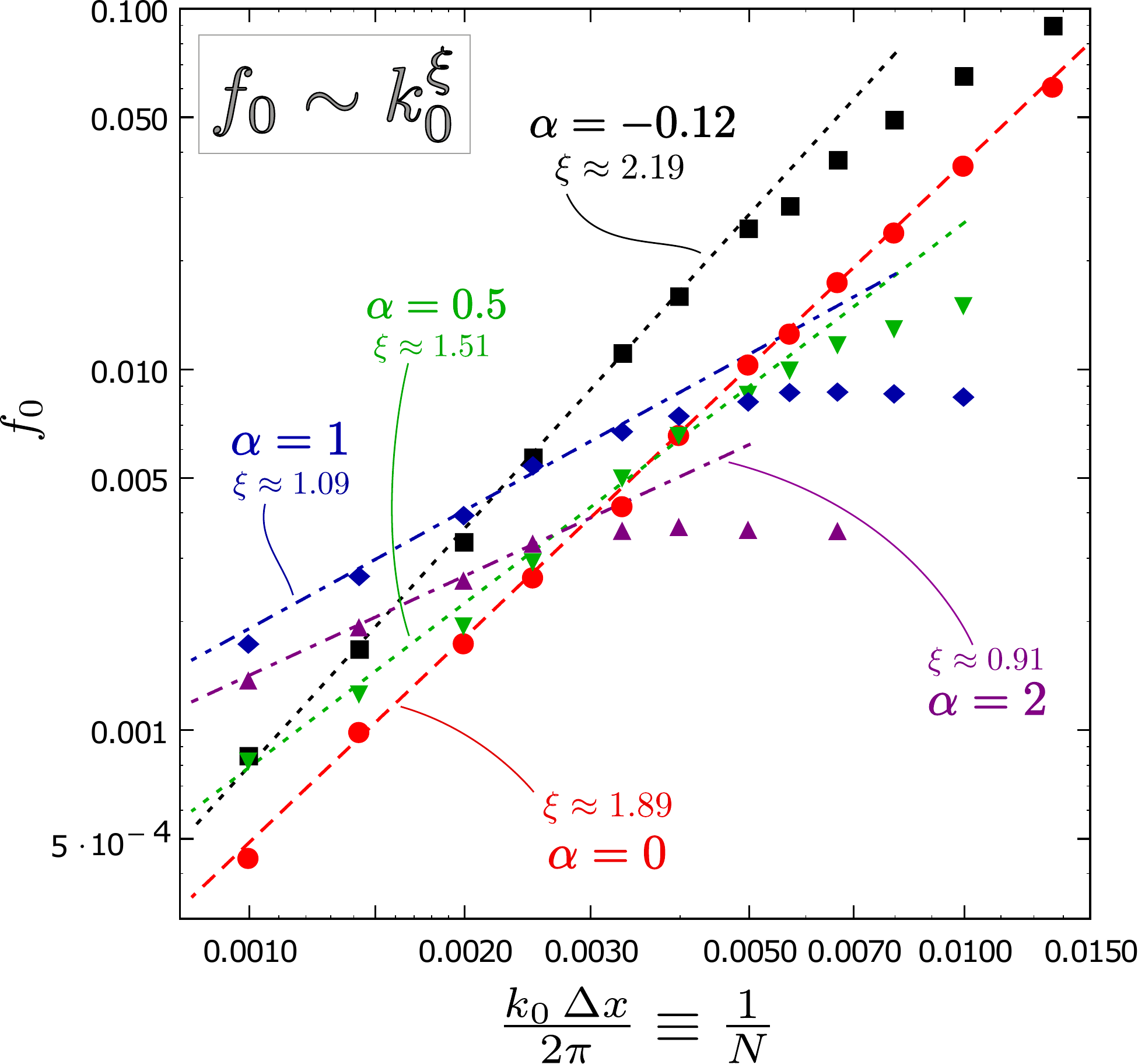}\par}
  \caption{(log-log scale) The lowest frequency $f_0$ dependence on the lowest wave number $k_0$ in the dynamics of 
  the surface roughness $w(t)$ for several parameter $\alpha$ values. Symbols: calculated values. Lines: power-law fits (with exponent $\xi$) of the results for small $k_0$.} 
  \label{fig:dispers-more}
\end{figure}

The PSDs of $w(t)$ for $-0.12\leq\alpha\leq 5$ can be fitted very well (see Fig.~\ref{fig:tispec_siz1000alphas}) by a generalized Lorentzian with an added constant (\ref{eq:gen-lorentzian_fit}) at different system sizes $N$ (except for some cases discussed in the following subsection). The exponent $\beta$ in the fit (\ref{eq:gen-lorentzian_fit}) increases monotonically from $\beta\approx 1.7$ for $\alpha=-0.12$ to $\beta\approx 3$ for $\alpha=5$.
From these fits at different system sizes $N$, the relations between the lowest frequencies $f_0$ in the dynamics and lowest wave numbers of spatial variations $k_0\propto N^{-1}$ are obtained (Fig.~\ref{fig:dispers-more}), as is done in Sec.~\ref{section:dynamics} for $\alpha=0$.

Fig.~\ref{fig:dispers-more} reveals how the spatio-temporal behaviour of evolving surfaces depend on parameter $\alpha$.

The relations $f_0(k_0)$ shown in Fig.~\ref{fig:dispers-more} indicate that for small $k_0$, the power-law behaviour $f_0\sim k_0^{\xi}$ observed in Fig.~\ref{fig:dispers000} for $\alpha=0$, also persists for $\alpha\neq 0$ with exponent $\xi$ decreasing with increasing $\alpha$: from $\xi\approx 2.2$ for $\alpha=-0.12$ to $\xi \approx 0.9$ for $\alpha=2$. However, for $\alpha\neq 0$, this power-law behaviour flattens out at larger values of $k_0$. 
For $\alpha=5$ (not shown in Fig.~\ref{fig:dispers-more}), the possible power-law behaviour is more difficult to determine, since the curve $f_0(k_0)$ appears flat almost through the whole range of $k_0$, except for only two points with smallest $k_0$ --- way less than enough to make conclusions.

One can interpret $f_0$ at some $k_0/ 2\pi=l_0^{-1}$ as the approximate rate of processes at the length scale $l_0$, or $n_0\equiv l_0/\Delta x$ in lattice units. Then the results displayed in Fig.~\ref{fig:dispers-more} imply that at smaller scales  --- say, $n_0<200$ ($k_0 \Delta x/ 2\pi>0.005$ in Fig.~\ref{fig:dispers-more}) --- the rate is monotonically decreasing with $\alpha$. On the other hand, for larger scales, this does not hold any more. For example, for $\alpha=0,\,0.5,\,1$ the relation of between $f_0$ and $\alpha$ reverses (becomes monotonically increasing) already at $n_0>300$. For large enough scales, $f_0$ should become monotonically increasing with $\alpha$ for all values, at least in $-0.12\leq\alpha\leq 2$, if the power-law trends $f_0(k_0)\propto k_0^\xi$ shown as straight lines in Fig.~\ref{fig:dispers-more} continue for even larger systems, $N>1000$.

\begin{figure}[ht!]
  {\centering\includegraphics[width=0.45\textwidth]{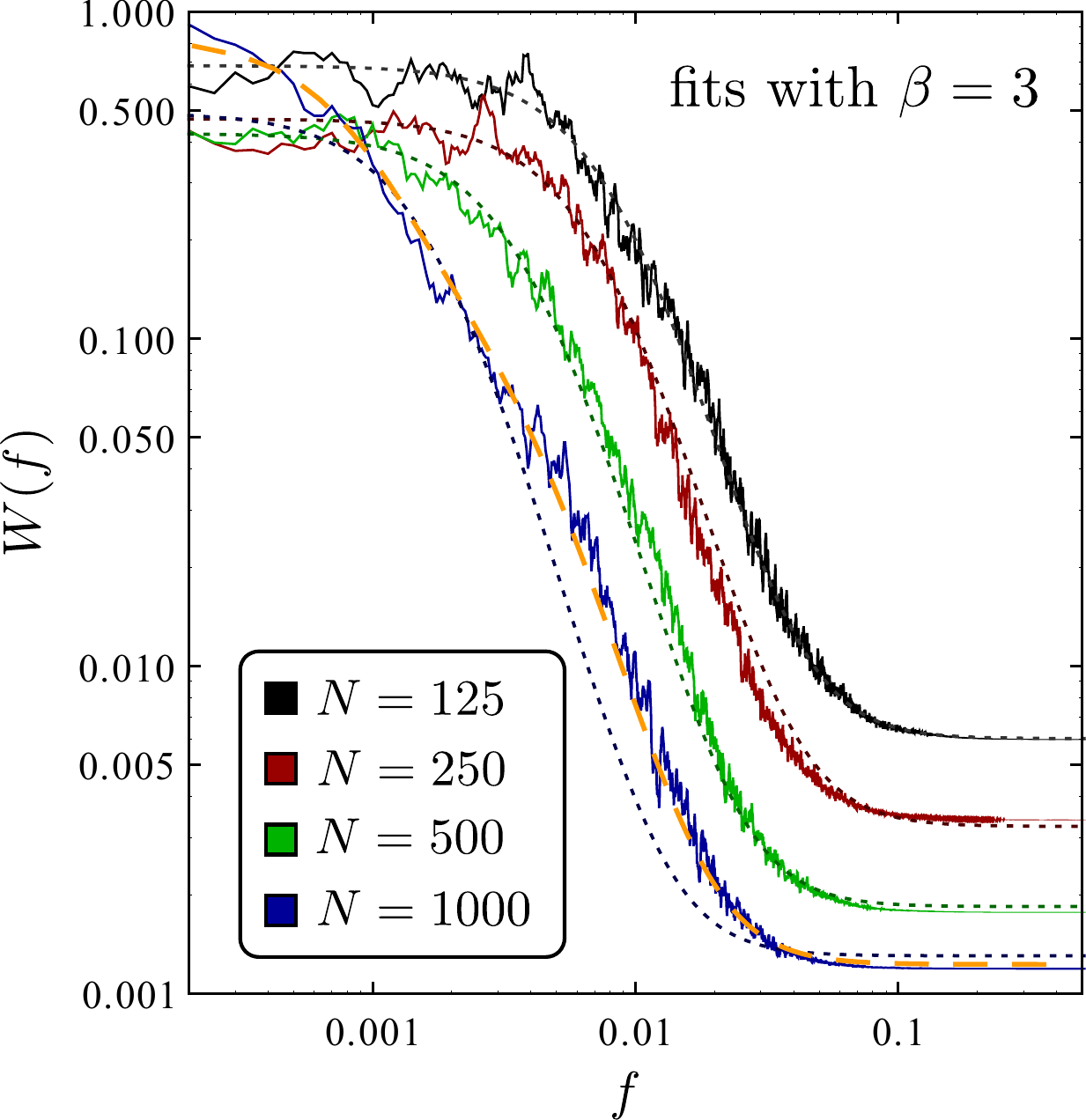}\par}
  \caption{The PSDs $W(f)$ of $w(t)$ from $t\in[2\cdot10^4,10^5)$ for $\alpha=1$ and different system sizes. Solid lines: the calculated PSDs. Dotted lines: single-generalized-Lorentzian (\ref{eq:gen-lorentzian_fit}) fits with exponent $\beta=3$. Long-dashed orange line: fit of the PSD at $N=1000$ by two generalized Lorentzians (\ref{eq:2lorfit}) with the same exponent $\beta=3$.} 
  \label{fig:tispec_alpha100sizes-both}
\end{figure}

\subsection{Fits by two generalized Lorentzians}

\begin{figure}[ht!]
  {\centering\includegraphics[width=0.45\textwidth]{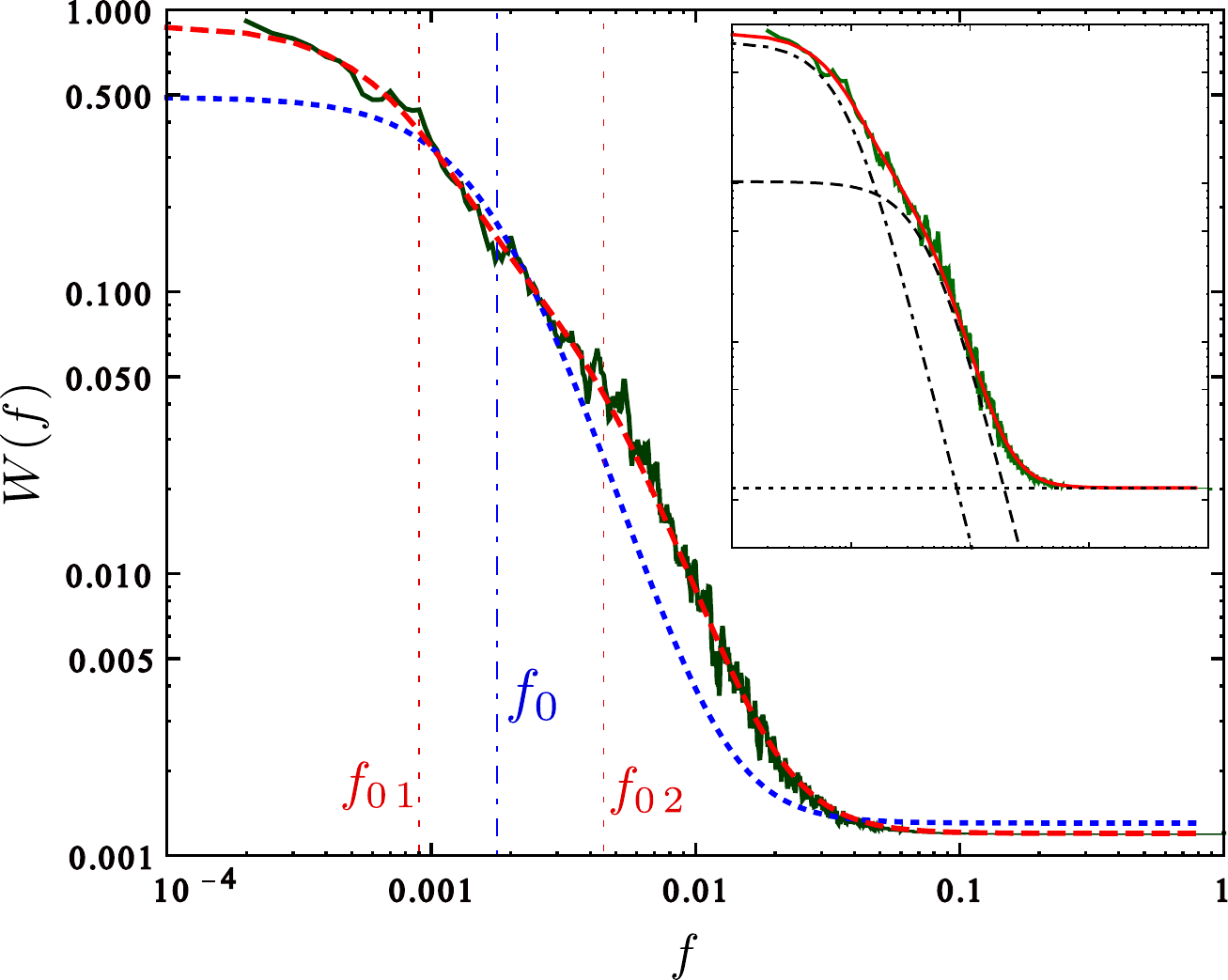}\par}
  \caption{(Log-log scale) The PSD of 
  the surface roughness $w(t)$ for $\alpha=1$ and $N=1000$ (solid dark green line) with fits by a single generalized Lorentzian (\ref{eq:gen-lorentzian_fit}) (dotted blue line) and by a two-generalized-Lorentzian fit (\ref{eq:2lorfit}) (dashed ref line),
c.f.~Fig.~\ref{fig:tispec_alpha100sizes-both}.  
   The vertical straight lines indicate the characteristic frequencies $f_0$, $f_{0\,1}$ and $f_{0\,2}$ of the fits. The inset shows the fit (\ref{eq:2lorfit}) decomposed into two Lorentzians and a constant.} 
  \label{fig:tispec100siz1000_specfit15_2lordeco-inset}
\end{figure}

The fits of the PSDs $W(f)$ by a generalized Lorentzian plus a constant (\ref{eq:gen-lorentzian_fit}) seem to be suitable for most cases investigated for $-0.12\leq \alpha \leq 5$ with system  sizes $100\leq N \leq 1000$. However, for $\alpha=0.5$ and $\alpha=1$, and system sizes $N\geq 700$, some larger deviations from the fits can be observed. 
For example, Fig.~\ref{fig:tispec_alpha100sizes-both} displays the apparent occurrence of a second hump in the PSD for $\alpha=1$ at $N=1000$ which renders the fit (\ref{eq:gen-lorentzian_fit}) less suitable, although at smaller $N$ it works very well (dotted lines in Fig.~\ref{fig:tispec_alpha100sizes-both}).
In these cases, however, the sum of two generalized Lorentzians and a constant with the same exponent $\beta$,
\begin{equation}
	W_{\mathrm{fit}}(f)=\frac{A_1}{(f_{0\,1}^2 + f^2)^{\beta/2}} + 
	\frac{A_2}{(f_{0\,2}^2 + f^2)^{\beta/2}} + B\,,
	\label{eq:2lorfit}
\end{equation}
fits the PSD almost perfectly (orange long-dashed line in Fig.~\ref{fig:tispec_alpha100sizes-both} and red dashed line in Fig.~\ref{fig:tispec100siz1000_specfit15_2lordeco-inset}).

As can be seen in Fig.~\ref{fig:tispec100siz1000_specfit15_2lordeco-inset}, the characteristic frequencies $f_{0\,1}$ and $f_{0\,2}$ of the two-generalized-Lorentzian fit (\ref{eq:2lorfit}) have the frequency $f_0$ of the original single-generalized-Lorentzian fit (\ref{eq:gen-lorentzian_fit}) between them, i.e., $f_{0\,1}<f_0<f_{0\,2}$. Moreover, the frequency $f_0$ seems to follow the power-law trend (blue diamonds and dash-dotted line in Fig.~\ref{fig:dispers-more}), even if the fit is not that good as for smaller $N$ values.

\begin{figure}[ht!]
  {\centering\includegraphics[width=0.45\textwidth]{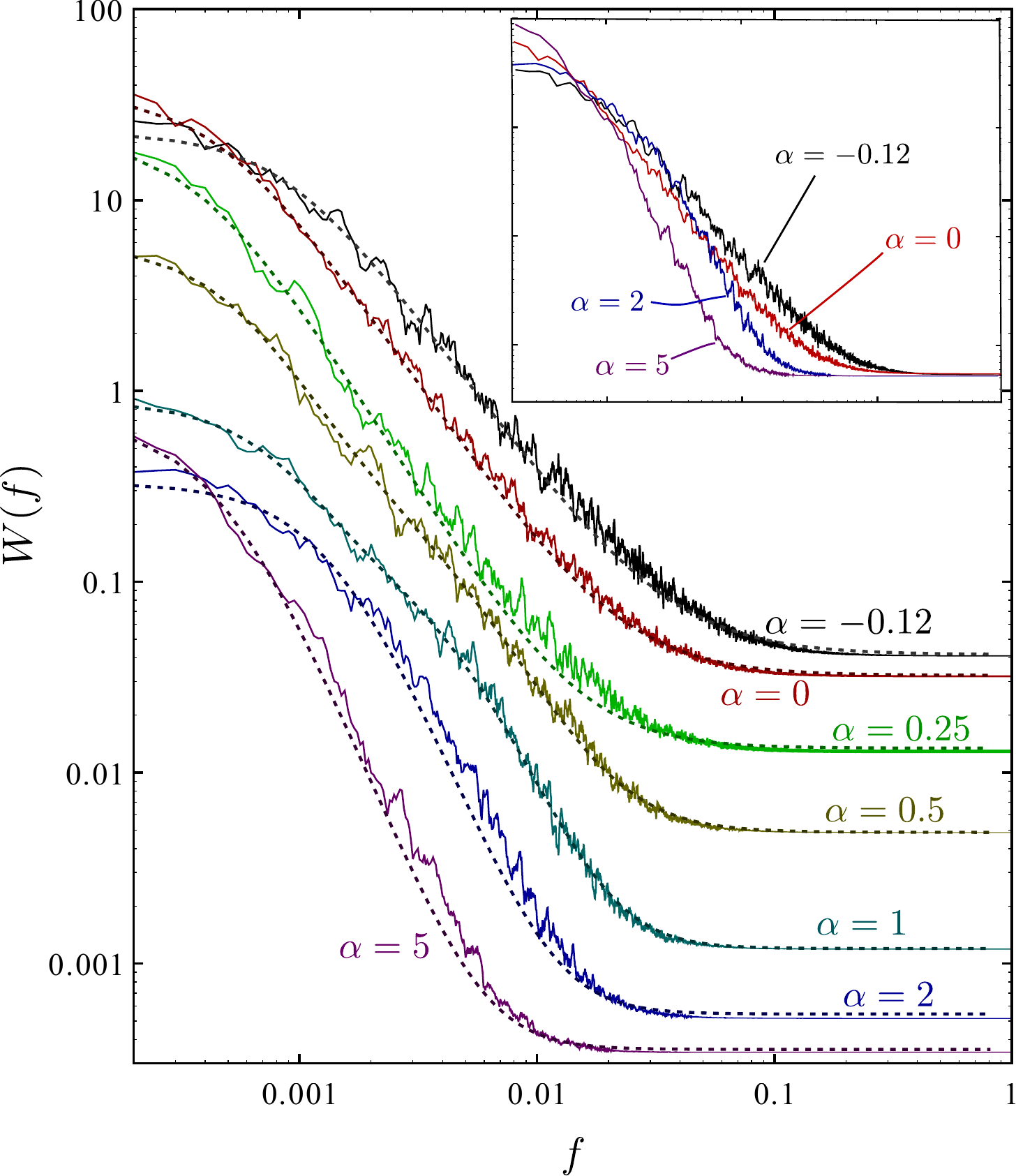}\par}
  \caption{Power-spectral densities (solid lines) of 
  the surface roughness $w(t)$ for a system size $N=1000$
  and different values of $\alpha$ with their fits (dotted lines): a two-generalized-Lorentzian fit (\ref{eq:2lorfit}) for $\alpha=0.5$ and $\alpha=1$, and a single generalized Lorentzian fit (\ref{eq:gen-lorentzian_fit}) for all other values. The inset shows some of the same PSDs normalized.} 
  \label{fig:tispec_siz1000alphas}
\end{figure}

Fig.~\ref{fig:tispec_siz1000alphas} displays the PSDs with their fits (\ref{eq:gen-lorentzian_fit}) and (\ref{eq:2lorfit}) for the whole parameter $\alpha$ range investigated at system size $N=1000$.

\section{Summary and outlook}
\label{sec:summary}

The results presented in this paper give some new insights into the complex spatio-temporal behaviour of surfaces produced by the two-dimensional generalized Kuramoto-Sivashinsky equation (\ref{eq:gkse}) and might be interesting to a broader circle of researchers working in the field of continuum systems with complex nonlinear dynamics. 

The scaling properties (\ref{wsquare_scaling}) of the saturated surface roughness indicate that additional large scale height variations of scale free character appear when the system size $N$ is increased. 
The dynamics of these slow height variations can be investigated by analysing the time series of the fluctuating surface roughness $w(t)$ where the occurrence of slow modes with increasing system size can also be observed (see Fig.~\ref{fig:tievo000sizes} and Fig.~\ref{fig:recplots000sizes}). 
This analysis shows that the resulting power-spectral densities (PSDs) can be expressed as the sum of a generalized Lorentzian and a constant, (\ref{eq:gen-lorentzian_fit}), or, in some cases, as two generalized Lorentzians (\ref{eq:2lorfit}), as shown in Figs.~\ref{fig:tispec000_fits}, \ref{fig:tispec_alpha100sizes-both} and \ref{fig:tispec_siz1000alphas}. 

The characteristic frequency $f_0$ obtained as a fit parameter corresponds to the smallest rate (largest time scale) that plays a role in the surface evolution. 
It can be attributed to the spatial mode of lowest wave number $k_0$ (which is inversely proportional to the system size) that can appear in the system. 
The dependence of this characteristic frequency on the system size gives the 'dispersion relation' $f_0(k_0)$ that connects spatial and temporal scales of surface dynamics. These relations have the power-law $f_0\sim k_0^{\xi}$ character (see Fig.~\ref{fig:dispers-more}) for large systems (small $k_0$), thus, suggesting that the underlying temporal behaviour is scale free. Also, the exponent $\xi$ is found to decrease with increasing value of parameter $\alpha$.
These results indicate, among other things, that although the characteristic time scale of dynamics on smaller scales decreases very strongly with increasing $\alpha$, on large enough scales, this relation is reversed, i.e., the evolution on large scales is slower for smaller $\alpha$.

The findings presented in this paper also raise some interesting questions 
for further research. 
For example, it is apparent from Fig.~\ref{fig:tievo_siz500alphas} and from the values of the PSD exponent $\beta$ that the character of surface roughness dynamics depends quite strongly on parameter $\alpha$. The question arises how temporal properties on various scales change with $\alpha$ and what are the statistical properties of the apparent bursts observed for larger values of $\alpha$.

The Lorentzian shape, $W(f)\sim (f_0^2+f^2)^{-1}$, of the PSD and relation $f_0\sim k_0^{\xi}$ with $\xi\approx 2$ for $\alpha\approx 0$ also suggests a possible analogy between the large-scale fluctuations of surface roughness and a diffusive process with the probability density Fourier transformed in space and time
\cite{van1992stochastic},
\begin{equation}
	\hat{P}(k_0, f)\propto\frac{k_0^2}{(k_0^2 D)^2 + f^2}\,,
	\nonumber
\end{equation}
where $D$ is the diffusion constant independent of the system size. This correspondence becomes apparent when $f_0=D k_0^2$ is substituted in (\ref{eq:lorentzian}).
Thus, perhaps the slow kinetics of the surface roughness might even be reproduced by a random walk of a particle in some external potential which is implied by the fact that the process $w(t)$ is bounded and, consequently, $k_0$ does not go to zero for systems of finite size.
For larger $\alpha$ values where the corresponding PSD exponent $\beta\approx 3$ and $\xi<2$ this process would then correspond to anomalous diffusion.
Moreover, the fact that, for some parameter values, one more generalized Lorentzian has to be added to the in order to fit the calculated PSD for large systems (see Figs.~\ref{fig:tispec_alpha100sizes-both} and \ref{fig:tispec100siz1000_specfit15_2lordeco-inset}) suggests the emergence of one more time scale, or perhaps, the whole interval of time scales. 
Any conclusive answers about both, the exact character and the occurrence mechanism, of this regime require more data obtained from simulations on even larger systems.
\bibliography{largescale}

\end{document}